\documentclass[symmetry,article,accept,pdftex,moreauthors]{Definitions/mdpi} 
\usepackage{url} %underline
\usepackage{ulem} %underline
\usepackage{amsmath}
\usepackage{amsfonts}
\usepackage{siunitx}
\graphicspath{{./Figures/}}

\usepackage{xargs}

\usepackage{empheq}

\usepackage{xspace}

\usepackage{todonotes}

\newcommand{\magn}[1]{\left|#1\right|}

\newcommand{\ket}[1]{\left|#1\right\rangle}

\newcommand{\gft}{\mathrm{GF}(2)}

\newcommand{\appref}[1]{Appendix~\ref{#1}}
\newcommand{\secref}[1]{\ref{#1}}

\newcommand{\tr}[1]{\mathrm{Tr}\left[ #1 \right]}

\newcommand{\stabs}{\mathcal{S}}

\DeclareMathOperator{\Tr}{Tr}

\firstpage{1} 
\pubvolume{14}
\issuenum{4}
\articlenumber{666}
\pubyear{2022}
\copyrightyear{2022}
\externaleditor{{Academic} Editor: Jim Freericks}
\datereceived{23 February 2022} 
\dateaccepted{22 March 2022} 
\datepublished{24 March 2022} 
\hreflink{https://doi.org/10.3390/\linebreak sym14040666} % If needed use \linebreak
\Title{Tunable Geometries in Sparse Clifford Circuits}

\TitleCitation{Tunable Geometries in Sparse Clifford Circuits}

\Author{Tomohiro Hashizume $^{1,}$*\orcidA{}, Sridevi Kuriyattil $^{1}$\orcidB{}, Andrew J. Daley
 $^{1}$\orcidC{} and Gregory Bentsen $^{2,}$*\orcidD{}}

\AuthorNames{Tomohiro Hashizume, and Sridevi Kuriyattil, and Andrew J. Daley, and Gregory Bentsen}

\AuthorCitation{Hashizume, T.; Kuriyattil, S.; Daley, A.J.; Bentsen, G.}
\address{%
$^{1}$ \quad Department of Physics and SUPA, University of Strathclyde, Glasgow G4 0NG, UK; kuriyattil-sridevi@strath.ac.uk (S.K.); andrew.daley@strath.ac.uk (A.J.D.) \\
$^{2}$ \quad Martin A. Fisher School of Physics, Brandeis University, Waltham, MA 02453, USA}

\corres{Correspondence: tomohiro.hashizume@strath.ac.uk (T.H.); gbentsen@brandeis.edu (G.B.)
}

\abstract{
    We investigate the emergence of different effective geometries in stochastic Clifford circuits with sparse coupling. By changing the probability distribution for choosing two-site gates as a function of distance, we generate sparse interactions that either decay or grow with distance as a function of a single tunable parameter. Tuning this parameter reveals three distinct regimes of geometry for the spreading of correlations and growth of entanglement in the system. We observe linear geometry for short-range interactions, treelike geometry on a sparse coupling graph for long-range interactions, and an intermediate fast scrambling regime at the crossover point between the linear and treelike geometries. This transition in geometry is revealed in calculations of the subsystem entanglement entropy and tripartite mutual information. We also study emergent lightcones that govern these effective geometries by teleporting a single qubit of information from an input qubit to an output qubit. These tools help to analyze distinct geometries arising in dynamics and correlation spreading in quantum many-body systems. 
}

\keyword{many-body entanglement; quantum circuits; lightcones; scrambling; long-range interactions; quantum teleportation} 

\begin{document}

% \listoftodos
% % Do not print any [non-inline] notes
% \makeatletter
% \renewcommand{\@todonotes@drawMarginNoteWithLine}{}
% \makeatother

% \newpage

% \tableofcontents
% \newpage
% \clearpage
% \pagenumbering{arabic}

%{{{ Main Text
%%%%%%%%%%%%%%%%%%%%%%%%%%%%%%%%%%%%%%%%%%
\section{Introduction}

% \sectabst{Greg: Motivate connection between entanglement and geometry, with connections to prospects in near-term experiments. Introduce Lieb--Robinson bounds and lightcones as probes of geometry.}

To describe entanglement in quantum many-body systems it is often beneficial to invoke concepts from geometry.
% To describe entanglement in a many-body quantum system it is often fruitful to invoke geometrical concepts such as areas, volumes, and minimal surfaces.
% In describing entanglement in many-body quantum systems it has often proved fruitful to invoke geometrical concepts such as minimal surfaces, areas, and volumes.
% In studying the dynamics of entanglement in many-body systems, it has often proven fruitful to view entanglement through the lens of geometrical concepts such as areas, volumes, and minimal surfaces.
A classic example is the distinction between `area-law' entanglement typically found in ground states of gapped systems versus the `volume-law' ground-state entanglement found in gapless systems or at quantum critical points \cite{Gioev2006EntanglementEntropy,Wolf2006Violation,hastings2006spectral,vidal2007entanglement,hastings2007area,Eisert2010_Area_Law_Review,
bianchiVolumelawEntanglementEntropy2021}. The geometrical notion of `area' naturally appears in this context because the entanglement in gapped systems is necessarily short-ranged, so that the entanglement entropy of a subregion $A$ is given by a minimal cut along the boundary $\partial A$ required to disentangle the region from its neighbors.

Similar connections between entanglement and geometry have also emerged in more recent studies of many-body dynamics. For example, it has become increasingly clear that entanglement dynamics in random circuits can be understood both heuristically and quantitatively in terms of domain walls and membranes \cite{Nahum2017,Nahum2018,skinnerMeasurementInducedPhaseTransitions2019,li2021statistical}. These minimal surfaces separate a subregion of interest from the rest of the system, similar to the minimal cuts described above.
% The AdS/CFT correspondence provides a direct link between entanglement in a quantum many-body system and the classical geometry of the bulk.
The AdS/CFT correspondence \cite{maldacena1999large,gubser1998gauge,witten1998anti,hartnoll2009lectures} provides another clear example of the close relationship between many-body entanglement and geometry,
% close relationship between entanglement and geometry in many-body quantum systems.
where an emergent `bulk' spacetime geometry is encoded into the entanglement patterns of a strongly-interacting quantum system located at the boundary of the spacetime \cite{swingle2012entanglement,Gubser_2017_p-adic_Ads,heydeman2017tensor_Ads}. The link between geometry and entanglement is made explicit via the Ryu--Takayanagi formula, which gives a geometric prescription for computing the entanglement entropy of the boundary theory in terms of a minimal surface anchored at the boundary and traversing through the bulk~\cite{ryu2006holographic,Ryu_2006_aspects_holo_ent,Hubey_2007_holo}. 

\begin{figure}[H]
    \centering
    \includegraphics[width=\columnwidth]{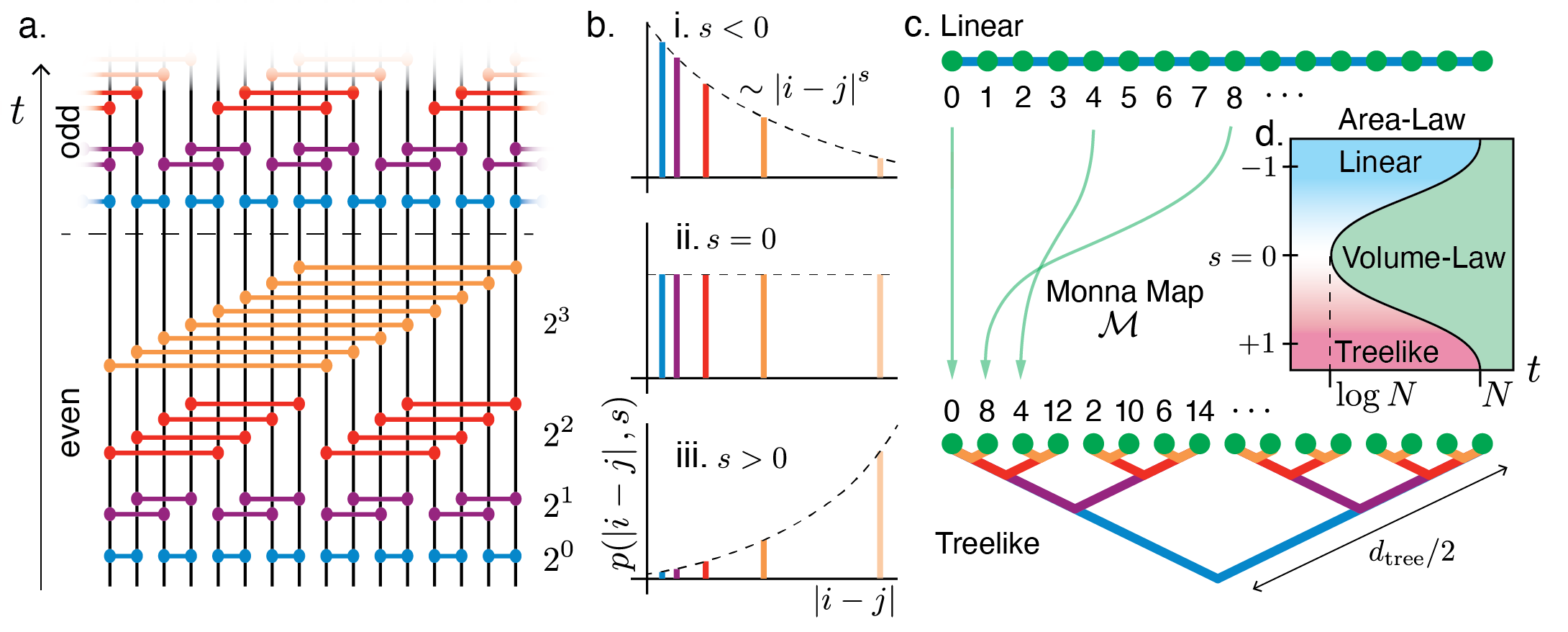}
    \caption{Sparse Clifford circuits yielding tunable geometries. Qubits $i,j$ arranged in a linear chain interact pairwise (\textbf{a}) if and only if they are separated by a power of two, $\magn{i-j} = 2^{m-1}$ for $m = 1,2,\ldots,\log_2 N$. To mimic continuously-tunable interactions in a discrete circuit, we arrange random two-qubit gates into even and odd bricklayer blocks, and randomly apply each gate $Q_{ij}$ with probability $p(\magn{i-j},s)$ given in Equation \eqref{eq:probcoupling}. A tunable power-law exponent $s$ (\textbf{b}) controls whether this distribution decays ($s < 0$, top) or grows ($s > 0$, bottom) with distance. The structure of entanglement generated by these circuits indicates linear (Euclidean) geometry when $s < 0$ and the treelike geometry of the $2$-adic numbers when $s > 0$ (\textbf{c}). Numerical studies of entanglement entropies as a function of time reveal three distinct regimes (\textbf{d}): a linear area-law regime (blue) when $s < 0$, a treelike area-law regime (red) when $s > 0$, and a volume-law regime (green) at late times. In the two local regimes $\magn{s} > 1$, volume-law entanglement builds up on a timescale $t_* \sim N$ that grows linearly with the system size, but near $s = 0$ fast scrambling dynamics generates volume-law entanglement on timescales as short as $t_* \sim \ln{N}$.}
    \label{fig:overview}
\end{figure}

At the same time, increasingly sophisticated experiments with cold neutral atoms, trapped ions, and superconducting qubits have demonstrated the ability to controllably engineer and probe complex patterns of many-body entanglement in the laboratory \cite{islam2015measuring,hucul2015modular,kaufman2016quantum,hosten2016measurement,Lukin_2019_entangl_probe,Tiff_2019_Probe_Renyi}. These capabilities raise an intriguing challenge:
% Can we engineer similar emergent geometries in qubit models that can be implemented in near-term experiments?
can we observe signatures of emergent geometry in the lab by engineering controlled patterns of entanglement in near-term cold atom experiments? First steps in this direction have recently been explored in pioneering experiments with nonlocally-interacting spins in a cavity system \cite{periwal2021programmable}, where spin--spin correlations were used to reconstruct the underlying geometry of the system. Similarly, numerical simulations \cite{bentsen2019treelike}, and field theory calculations \cite{gubser2018continuum,gubser2019mixed} have been used to study sparse systems of this kind and to study the effective geometries that emerge from the dynamics, but these methods only allowed access to dynamics with weak interactions or at short times. Here we extend these ideas by numerically studying entanglement dynamics in sparse Clifford circuits, which enable numerical access to strong interactions at large system sizes for a special class of many-body dynamics. The Clifford circuits we study feature sparse, tunable long-range interactions (Figure \ref{fig:overview}a,b), of the type that can be engineered in multi-drive cavities \cite{bentsen2019treelike,periwal2021programmable} or Rydberg arrays with tweezer-assisted shuffling \cite{hashizumeDeterministicFastScrambling2021}.
% computing entanglement entropies in strongly-scrambling systems featuring tunable sparse interactions, and provide numerical and analytical evidence for three distinct types of geometry that emerge in such systems. Specifically, we explore the emergence of linear, treelike, and black-hole-like geometries generated by the patterns of entanglement found in quantum circuits.

To directly access the geometry of these quantum circuits we study their lightcones, which govern the spreading of quantum information through the system. In 1 + 1d systems with short-ranged interactions, Lieb--Robinson bounds yield a linear lightcone \cite{lieb1972finite} which limits the propagation of quantum information. Outside of the many-body lightcone, correlation functions decay exponentially; as a result, initially localized information is prohibited from spreading through the entire system earlier than a time $t \sim \mathcal{O}(N)$ that grows linearly with the system size. Similar Lieb--Robinson bounds may be derived for more generic quantum systems \cite{hastings2010locality,roberts2016lieb,elseImprovedLiebRobinsonBound2018}, including systems with long-range and sparse interactions \cite{elseImprovedLiebRobinsonBound2018,bentsen2019fast,tranLiebRobinsonLightCone2021}. Inside the many-body lightcone, strongly-interacting chaotic systems tend to rapidly \emph{scramble} quantum information, encoding it into complicated patterns of entanglement between all qubits lying inside the lightcone. 
% We therefore expect to be able to map out the local lighcone of a generic chaotic quantum system by studying the entanglement entropies $S_A$ of geometrically contiguous subregions of output qubits.
When the lightcone eventually spreads to all $N$ qubits, quantum information is delocalized across the entire many-body system and the system becomes Page scrambled \cite{Page1993}, with volume-law entanglement at all scales. In systems defined by a local geometry, the Lieb--Robinson lightcone always prevents such system-wide scrambling any sooner than a time of order $t_* \sim N$ scaling at least linearly with system size. However, in systems with long-range and sparse interactions the local lightcone collapses, allowing for \emph{fast scrambling}, where system-wide entanglement builds up on timescales as short as $t_* \sim \ln{N}$ \cite{sekino2008}.

% just such a lightcone collapse in a model of sparsely-interacting qubits

% Entanglement entropies $S_A$ for subregions of qubits therefore tend to display volume-law entanglement $S_A \sim \magn{A}$ for regions $A$ lying inside the lightcone but display area-law entanglement $S_A \sim \mathrm{const.}$ for regions much larger than the lightcone. Entanglement entropies $S_A$ as a function of subregion size and time therefore provide a sharp probe of the many-body lightcone, rooted in the system's local geometry.

% In particular, many-body entanglement can only build up significantly inside the system's lightcone. For strongly-interacting chaotic dynamics, we therefore generically expect volume-law entanglement inside the many-body lightcone and area-law entanglement for regions that are much larger than the 

% Even strongly-interacting chaotic systems, which tend to build up entanglement at the fastest rate possible, are unable to 

% These linear lightcones exist precisely because the underlying geometry is linear and 1+1d. These Lieb--Robinson bounds and their corresponding lightcones clearly illuminate the underlying geometry, 

Here we study the emergence and collapse of tunable local lightcones in a model of sparsely-coupled qubits. Our circuits feature sparse interactions, where pairs of qubits in a 1d chain are coupled if and only if they are separated by a power of 2 as illustrated in Figure \ref{fig:overview}a. A tunable real parameter $s$ controls whether these interactions decay ($s < 0$) or grow ($s > 0$) with distance as illustrated in Figure \ref{fig:overview}b and leads to three distinct effective geometries governed by different emergent lightcones as shown in Figure \ref{fig:overview}c.
% Here we search for emergent lightcones in our sparse Clifford circuits as a signature of the underlying geometry. We find that this geometry changes dramatically as a function of a tunable parameter $s$ that controls how the sparse couplings grow or decay with distance as depicted in \figref{fig:overview}b.
For large negative $s < 0$ we find a conventional linear lightcone similar to those found in short-ranged systems, indicating the conventional linear (Euclidean) geometry expected for interactions that are local in space (Figure \ref{fig:overview}d, blue). At late times $t \sim N$, after the linear lightcone has spread through the entire system, we find volume-law entanglement throughout the circuit (Figure \ref{fig:overview}d, green).
By contrast, for positive $s > 0$ we find an emergent lightcone governed by the treelike (Ultrametric) geometry of the 2-adic numbers \cite{rammalUltrametricityPhysicists1986,gubser2018continuum,bentsen2019treelike}. This geometry arises naturally in this case, because long-ranged interactions between spins that are separated by the highest available power of two in the given system size dominate the dynamics. It then makes sense to consider the treelike reordering of spins in which the spins that interact most strongly are next to each other, as shown in Figure \ref{fig:overview}c. In this treelike geometry the distance between a pair of qubits $i,j$ is measured by the depth of the shortest binary tree connecting the qubits as illustrated in Figure \ref{fig:overview}c.
% is defined by a treelike metric where distances between qubits are measured that can be achieved via performing Monna mapping to the linear arrangement of the qubits (\figref{fig:overview}c). The geometry in large positive $s>0$ limit are studied closely on spin-chains \cite{dysonExistencePhasetransitionOnedimensional1969} 
% and free fermions \cite{gubser2018continuum} along the context of an ultrametric geometry \cite{rammalUltrametricityPhysicists1986}.
% Here we find that a treelike lightcone governs the flow of quantum information when $s > 0$, whereas a conventional linear lightcone applies when $s < 0$ .
Similar to the short-ranged case, this treelike lightcone prohibits quantum information scrambling before a time $t_* \sim N^a$ that grows polynomially with the system size (Figure \ref{fig:overview}d, red). Near the crossover point $s = 0$, however, both the linear and treelike lightcones collapse, allowing for fast scrambling dynamics capable of maximally scrambling quantum information on a timescale $t_* \sim \ln{N}$ that grows logarithmically with the system size.

In the following sections we characterize these emergent lightcones and their corresponding geometries---linear, treelike, and fast scrambling---by numerically computing entanglement entropies in a family of Clifford circuits with sparse nonlocal interactions with tunable parameter $s$. In Section \ref{sec:models}, we introduce the sparse Clifford circuits studied in this work. In Section \ref{sec:arealawvolumelaw}, we study entanglement entropies $S_A$ of contiguous regions $A$ of output qubits prior to the scrambling time, and show that the area-law or volume-law scaling of the entropy in these regions at early times already allows us to extract some information about the geometry. In Section \ref{sec:trimutualinfo}, we study the scrambling time $t_*$ in different regimes of $s$ as diagnosed by the negativity of tripartite mutual information $I(A:B:C)$ between three equal-sized contiguous regions, and show that the scrambling time scales polynomially with system size $t_* \sim N^a$ in both the linear and treelike regimes, but collapses to a logarithmic scaling $t_* \sim \ln{N}$ near the fast-scrambling limit $s = 0$. In Section \ref{sec:telelightcone}, we directly probe the geometry of the circuit by mapping the system's many-body lightcone, which we extract from the fidelity $I(i;j|\mathrm{m})$ of teleporting a quantum state from a fixed input qubit $i$ to a desired output qubit $j$. Finally, we summarize our findings in Section \ref{sec:discussion} and conclude with some possible directions for future work.

\section{Sparse Clifford Circuits}
\label{sec:models}

% \sectabst{Greg: Introduce Clifford circuits with sparse long-range interactions. Exponent $s$ mimicked by placing gates with probability $p(s)$.}

In this paper, we study Clifford circuit models featuring sparse nonlocal interactions, where spin-1/2 qubits residing at sites $i,j = 0,\ldots,N-1$ of a 1d chain interact if and only if they are separated by an integer power of 2: $\magn{i-j} = 2^{m-1}$ for $m = 1,\ldots,\log_2 N$ \mbox{(Figure \ref{fig:overview}a)}. A tunable parameter $s$ controls how these couplings either decay ($s < 0$) or grow ($s > 0$) with distance as illustrated in Figure \ref{fig:overview}b. Recent analytical \cite{gubser2018continuum,gubser2019mixed}, numerical \cite{bentsen2019treelike,hashizumeDeterministicFastScrambling2021}, and experimental \cite{periwal2021programmable} work has demonstrated that sparse nonlocal models of this type exhibit a transition in geometry as a function of the exponent $s$, yielding a linear (Euclidean) geometry for $s < 0$, a treelike (Ultrametric) geometry for $s > 0$, and a fast-scrambling geometry at the midpoint $s = 0$ between the two. These previous analyses~\cite{gubser2018continuum,gubser2019mixed,bentsen2019treelike,hashizumeDeterministicFastScrambling2021,periwal2021programmable} have argued that the flow of quantum information in such systems is governed by a linear lightcone when $s < 0$ and by a treelike lightcone when $s > 0$, and that both of these lightcones break down near $s = 0$, allowing for fast scrambling dynamics.

Here we study the emergence of similar lightcones in analogous stochastic Clifford circuits. We use Clifford circuits in particular because they can be efficiently classically simulated \cite{gottesman1998heisenberg,aaronson2004improved} (\appref{app:stabilizers}), while also capturing many-body entanglement. The dynamics of Clifford circuits closely mirror the dynamics of general random unitary circuits \cite{Nahum2018,liMeasurementdrivenEntanglementTransition2019} because Clifford circuits are unitary 2-designs \cite{cleveNearlinearConstructionsExact2016}.
Our circuits are composed of random two-qubit Clifford gates $Q_{ij} = Q_{ji}$, where the exponent $s$ controls the probability $p(\magn{i-j},s)$ of placing a gate $Q_{ij}$ between qubits $i,j$ during each timestep $\delta t$. In the sparse Clifford circuit, the gates are arranged in a nonlocal bricklayer pattern illustrated in Figure \ref{fig:overview}a, with interaction layers stacked into an alternating sequence of `even' and `odd' blocks. During the `even' block, a gate $Q_{ij}$ is placed between qubits $i < j$ with probability $p(\magn{i-j},s)$ if and only if $\textrm{mod}(\lfloor i / 2^{m-1} \rfloor,2) = 0$. 
During the subsequent \emph{odd} block, gates are placed according to the same rules but with the odd-bricklayer condition $\textrm{mod}(\lfloor i / 2^{m-1} \rfloor,2) = 1$~\cite{hashizume2021measurement}.
Throughout the paper a single timestep $\delta t$ corresponds to applying a single even block followed by a single odd block.
% When $s = 0$ each of these gates is placed with equal probability $p(\magn{i-j},0) = 1$, but for arbitrary $s$ we modulate 
The probability of placing each gate $Q_{ij}$ is modulated according to the sparse probability distribution 
\begin{equation}
    p(\magn{i-j},s) = \begin{cases} 
      J_s \magn{i-j}^s & \mathrm{when}\, \magn{i-j} = 2^{m-1} \\
      0 & \mathrm{otherwise}
   \end{cases}
   \label{eq:probcoupling}
\end{equation}
which mimics the sparse coupling matrix $J(\magn{i-j},s)$ used in analogous continuous-time models \cite{bentsen2019treelike}. 
The normalization factor $J_s$ ensures that one gate, on average, is applied per site during each timestep $\delta t$ (see \appref{app:normalization}). Thus, short-distance gates dominate the Clifford circuit when $s < 0$ while long-distance gates dominate when $s > 0$ (Figure \ref{fig:overview}b top and bottom). At the midpoint $s = 0$, gates at all length scales are equally probable in the circuit (Figure \ref{fig:overview}b middle).
Here and throughout the paper, we impose periodic boundary conditions, such that the linear distance $\magn{i-j}$ between a pair of sites is the smaller of $\mathrm{abs}(i-j)$ and $N-\mathrm{abs}(i-j)$.
% The sparse probability distribution $p(\magn{i-j},s)$ plays a role analogous to the sparse coupling matrix [$J(\magn{i-j},s)$] used in [earlier works].

\section{Probing Emergent Geometry with Entanglement Entropy}
\label{sec:arealawvolumelaw}

% \sectabst{Sridevi: Demonstrate linear, treelike, and fast-scrambling geometries by studying entanglement entropies $S_A$ as a function of region size $\magn{A}$ and time $t$. Additionally, introduce Monna map $\mathcal{M}$ and entanglement entropy $S_A$ in this section.}

\begin{figure}[H]
    \centering
    \includegraphics[width=\columnwidth]{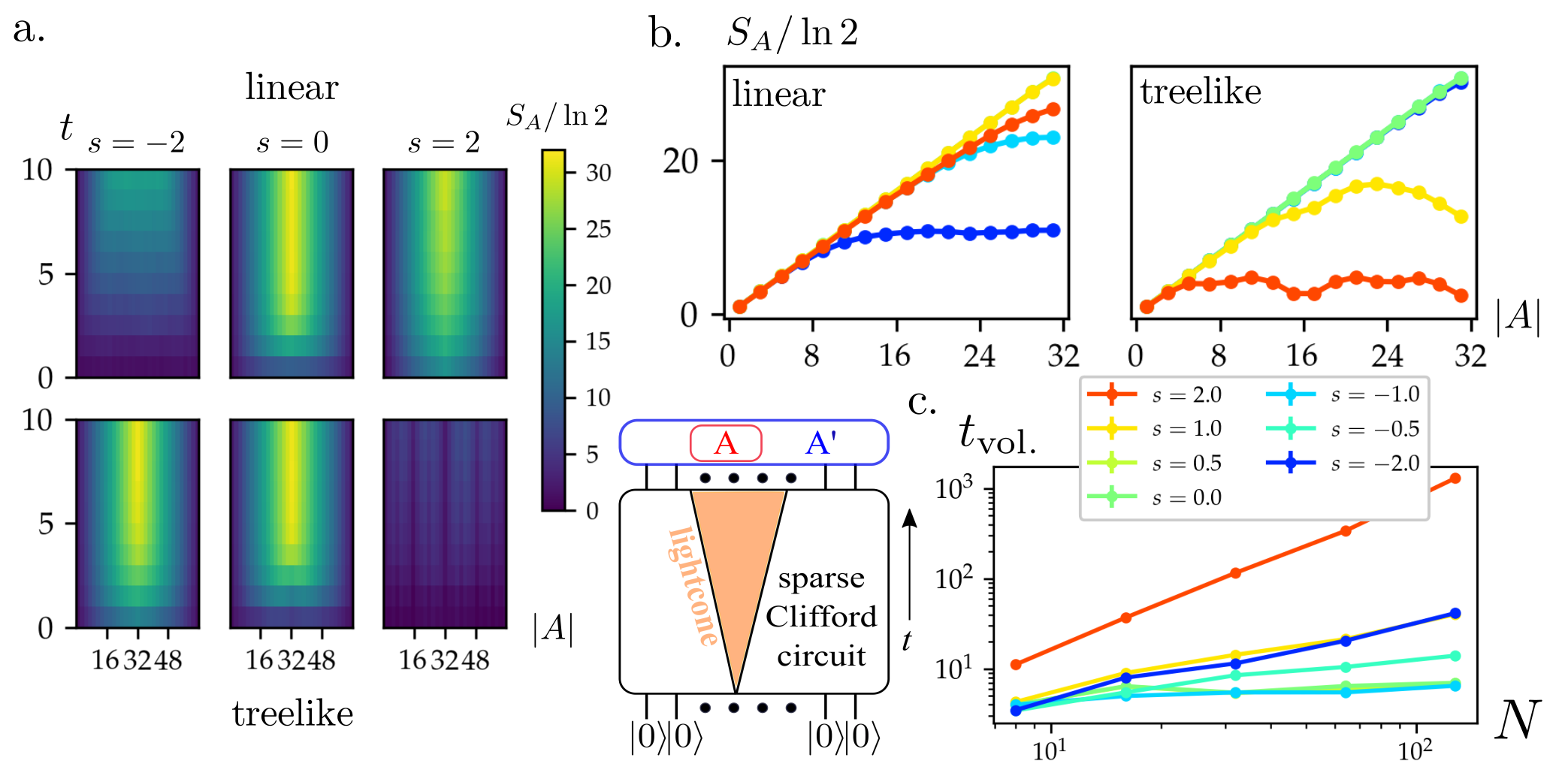}
    \caption{
       Probing tunable geometry using entanglement entropy.
       (\textbf{a}) Entanglement growth in contiguous subregions $A$ in a system of $N=64$ qubits initialized in the product state $\ket{0}^{\bigotimes N}$, plotted as a function of subregion size $\magn{A}$ and time $t$ in linear (top) and treelike (bottom) geometries. 
       (\textbf{b}) A time-slice at $t=6$ from the color-map illustrates the area-law scaling and volume-law scaling for 
       linear (left) and treelike (right) geometries. When $s < 0$, linearly contiguous subregions $A$ exhibit area-law entanglement at short times, while other subregions are trivially volume-law entangled.
    When $s > 0$, treelike subregions have area-law entanglement at short times, while other regions are volume-law entangled. 
 (\textbf{c}) The timescale $t_{\mathrm{vol.}}$ required to reach volume-law entanglement grows polynomially with system size $N=2^{3},\cdots,2^{7}$ when $\magn{s} > 1$, but is much shorter near $s = 0$, consistent with a logarithmic scrambling time $t_{\mathrm{vol.}}  \sim \ln{N}$. Error bars are shown or are smaller than data points. Data points are obtained by averaging over $10^{2}$ trajectories.}
    \label{fig:areavolumelaw}
\end{figure}

We begin characterizing the geometry generated by the circuit by examining the pattern of entanglement present in various subregions $A$ of the output qubits as illustrated in Figure \ref{fig:areavolumelaw}a.
% which serves as one probe of the underlying geometry generated during time evolution.
Because the circuit is composed entirely of Clifford gates, we may completely characterize the entanglement in the system at any time $t$ by computing Renyi entropies $S_A \equiv S_A^{(2)}=-\ln{\Tr[\rho_{A}^{2}]}$ of subregions $A$ \cite{renyi1961measures,aaronson2004improved}. In this section, we study Renyi entropies of \emph{contiguous} subregions $A$, where the meaning of the word `contiguous' depends on the geometry implied by the interactions. % As the circuit is entirely composed of Clifford gates, the Renyi entropy $S_A^{(2)}$ is always equal to the von-Neumann entropy $S_A = \Tr[\rho_A \ln{\rho_A}]$, and hence we may use $S_A^{(2)}$ to characterize the entanglement of the system.
A linearly-contiguous region $A$ has the property that any bipartition $A = A_1 \bigcup A_2$ is contiguous with respect to the Euclidean metric: for every $i \in A_1$, there exists a $j \in A_2$ such that $\magn{i-j} = 1$.
When $s < 0$, we shall find that the entanglement generated by the circuit is organized into linearly-contiguous subregions. %with contiguity defined by the usual linear (Euclidean) metric $\magn{i-j}$.

By contrast, in the limit $s > 0$, we shall find that the entanglement in the circuit is organized into treelike-contiguous regions defined by the treelike ($2$-adic) metric $\magn{i-j}_{2} = 2^{d_{tree}(i,j)/2}/N$, where $d_{tree}(i,j)$ is the number of edges required to connect sites $i$ and $j$ in the regular binary tree shown at the bottom of Figure \ref{fig:overview}c (see \appref{app:monnamap} for a more comprehensive discussion of the $2$-adic metric $\magn{i-j}_2$). A treelike-contiguous region $A$ is defined by the property that any bipartition $A = A_1 \bigcup A_2$ is contiguous with respect to the treelike ($2$-adic) metric: for every $i \in A_1$, there exists a $j \in A_2$ such that $\magn{i-j}_2 = 2/N$ for $N$ a power of 2.
One can obtain the sites $i,j$ in the treelike ordering by rearranging the qubits by the discrete Monna map $\mathcal{M}$, which reverses the digits of the argument when written in base 2 (c.f. \appref{app:monnamap}).
%always within a minimal 2-adic distance $\magn{i-j}_2 = 2/N$ of some other site $j \in A$. For any finite system of qubits $i = 0,1,\ldots,N-1$, one obtains the treelike ordering by rearranging the qubits by the discrete Monna map $\mathcal{M}$, which reverses the digits of the argument when written in base 2 (c.f. \appref{app:monnamap}).
For example, in a system with $N=8$ qubits, site $1$ is mapped to site $\mathcal{M}(1)=4$ under the Monna map because $\mathcal{M}(001_{2})=100_{2}$ when the site numbers are written in binary.
\vspace{-6pt}

\textls[-15]{Simply relabeling the sites under the Monna map $\mathcal{M}$ is not enough to constitute a distinct geometry, but the treelike structure of the 2-adic metric $\magn{i-j}_2$ endows this arrangement with a notion of distance that is entirely different from the conventional metric $\magn{i-j}$ in linear (Euclidean) geometry. Specifically, while both measures of distance satisfy the usual mathematical axioms required for a metric, the 2-adic measure is \emph{ultrametric}, meaning that it satisfies a much stronger form of the triangle inequality $\magn{i-j}_2 \leq \mathrm{max}(\magn{i-k}_2,\magn{j-k}_2)$ for all $i,j,k$. Additionally, while both geometries are translation-invariant \mbox{$(i,j) \rightarrow (i+\ell,j+\ell) \mod N$}, the treelike geometry is also invariant under a much larger number of nested permutation symmetries which exchange the left and right halves of any subtree. In particular, by consecutively applying exchange permutations at each level of the tree in Figure \ref{fig:overview}c we obtain translational invariance in the treelike geometry, i.e., $d_{tree}(i+1,j+1)=d_{tree}(i,j)$~\cite{huang_general_2021,heydeman2017tensor_Ads,stoica_building_2021,gubser_edge_2017,Gubser_2017_p-adic_Ads,bentsen2020tunable}.}

Inspired by the growth of entanglement in lattice systems, we expect typical contiguous subregions $A$ of output qubits to be volume-law entangled $S_A \sim \magn{A}$ when they are small enough to lie entirely inside the system's many-body lightcone, but to cross over to area-law entanglement $S_A \sim \mathrm{const.}$ when the region becomes much larger than the extent of the lightcone at a given time, as illustrated in Figure \ref{fig:areavolumelaw}a. On the contrary, geometrically disjoint subregions---for example, arbitrary subsets of $k \ll N$ qubits chosen at random from anywhere in the system---typically exhibit volume-law entanglement after only a single layer of gates, which originates trivially from the short-range entanglement in the system. This distinction between area-law and volume-law scaling in the entanglement entropy therefore provides a simple and sharp test of geometrical contiguity. Specifically, contiguous subregions $A$ should exhibit area-law entanglement scaling at short times, whereas non-contiguous subregions should exhibit volume-law scaling.

% We expect these statements to be true in both regimes $s < 0$ and $s > 0$.
We therefore proceed to use the scaling of the Renyi entropy with subregion $\magn{A}$ as a quantitative probe of the system's geometry.
This is clearly illustrated in Figure \ref{fig:areavolumelaw}, where we analyze the growth of entanglement in the sparse Clifford circuit acting on qubits initialized in a product state $\ket{0}^{\bigotimes N}$, where $\ket{0}$ is the ground state of the Pauli-$Z$ operator. 
%a product state of spins polarized along $\hat{\textbf{z}}$.  
The colour shading in Figure \ref{fig:areavolumelaw}a shows the entanglement entropy $S_A$ of contiguous subregions $A$ as a function of their size $\magn{A}$ and time $t$ with either linearly-contiguous bipartitions (top) or treelike-contiguous bipartitions (bottom). Taking a time-slice $t = 6$ from Figure \ref{fig:areavolumelaw}a, we find area-law scaling in the linear geometry at negative values of $s < 0$ (Figure \ref{fig:areavolumelaw}b, left), whereas other choices of geometry yield volume-law entanglement. Similarly, at positive values of $s > 0$ we find area-law scaling in the treelike geometry and volume-law entanglement for other choices of geometry (Figure \ref{fig:areavolumelaw}b, right). The non-monotonic features appearing at early times in Figure \ref{fig:areavolumelaw}b are a direct result of the treelike geometric structure present in the circuit at large $s > 0$. For instance, a treelike region consisting of precisely half the qubits $\magn{A} = N/2$ has especially low entropy at early times coming from the weakest long-range interactions in the deepest parts of the tree (blue couplings in Figure \ref{fig:overview}c). By contrast, smaller treelike regions $\magn{A} < N/2$ will have additional entropy coming from couplings higher in the tree (purple, red, orange in Figure \ref{fig:overview}c). Together, these observations confirm that the entanglement entropy has a volume-law scaling for typical regions $A$ and an area-law scaling $S_A \sim O(1)$ when the regions are chosen to be either linearly- or treelike-contiguous depending on the value of $s$.

Finally, we estimate the timescale $t_{\mathrm{vol.}}$ required for extensive contiguous regions $\magn{A} = N/2$ to saturate to volume-law entanglement. 
In the linear regime $s < 0$, we find that this timescale grows linearly in the system size $t_{\mathrm{vol.}} \sim N$.
In the treelike regime $s > 0$, on the other hand, we find that this timescale grows polynomially in the system size 
$t_{\mathrm{vol.}} \sim N^{|s|}$ (\appref{app:Linear_Fits}).
This is expected because the local geometry in each case inhibits the spreading of quantum information through the system, and an extensive subregion $A$ cannot become volume-law entangled until the lightcone of a single qubit has spread to of order $\magn{A}$ qubits \cite{tranLiebRobinsonLightCone2021}. By contrast, near the crossover point $s = 0$ the circuit achieves volume-law entanglement on much shorter timescales consistent with logarithmic scaling $t_{\mathrm{vol.}} \sim \ln{N}$, which indicates fast scrambling dynamics \cite{sekino2008,lashkari2011towards,yao2016interferometric,bentsen2019treelike,marino2019cavity,bentsen2019fast,li2020fast,Belyansky2020,hashizumeDeterministicFastScrambling2021}.
% The entanglement entropy $S_A$ of a subregion size $A=N/2$ saturates to a maximum value after $t_{*}$ interaction layers.
% Depending on whether the spins are partitioned either according to their physical position or treelike ordering, the saturation time $t_{*}$ changes. The saturation time for large positive $s$ is high for large system size $N$, because 

% We are particularly interested in whether regions $A$ exhibit `area-law' versus `volume-law' entanglement in which $S_A$ is largely independent of the r Prior to the scrambling time -- when 

\section{Scrambling and Negativity of Tripartite Mutual Information}
\label{sec:trimutualinfo}
% \sectabst{Tom: Find the scrambling time $t_*$ in linear, treelike, and fast-scrambling regimes by studying negativity of tripartite mutual information $I_3(A:B:C)$ between three contiguous regions of size $\magn{A} = N/4$.}

In the previous section we observed that local lightcones in the linear and treelike regimes prevent the buildup of system-wide entanglement until times of order $t \sim \mathcal{O}(N)$.
We can more precisely study the growth of these correlations by considering the scrambling of quantum information as diagnosed by the tripartite mutual information
\begin{align}
   I(A:B:C)=I(A;B)+I(A;C)-I(A;BC)
\end{align}
\textls[-11]{of three geometrically contiguous subregions $A,B,C$ of equal size $N/4$, where $I(A;B)=S_A+S_B-S_{AB}$ is the mutual information between regions $A$ and $B$. The tripartite mutual information vanishes when the regions are uncorrelated, but becomes negative \mbox{$I(A:B:C) < 0$}} as entanglement builds up across the system \cite{haydenHolographicMutualInformation2013,hosur2016chaos,gullans2020dynamical}. Negativity of the tripartite mutual information for three extensive regions $A,B,C$ therefore serves as a natural measure of the degree to which nonlocal correlations have built up across the entire system.

% The as qubits scramble, the negativity of the quantity $I(A:B:C)$ increases, implying that information is being shared across $A,B,C$
% rather than stored locally in the regions $A$, $B$, or $C$.

\begin{figure}[H]
   \centering
   \includegraphics[width=1.0\textwidth]{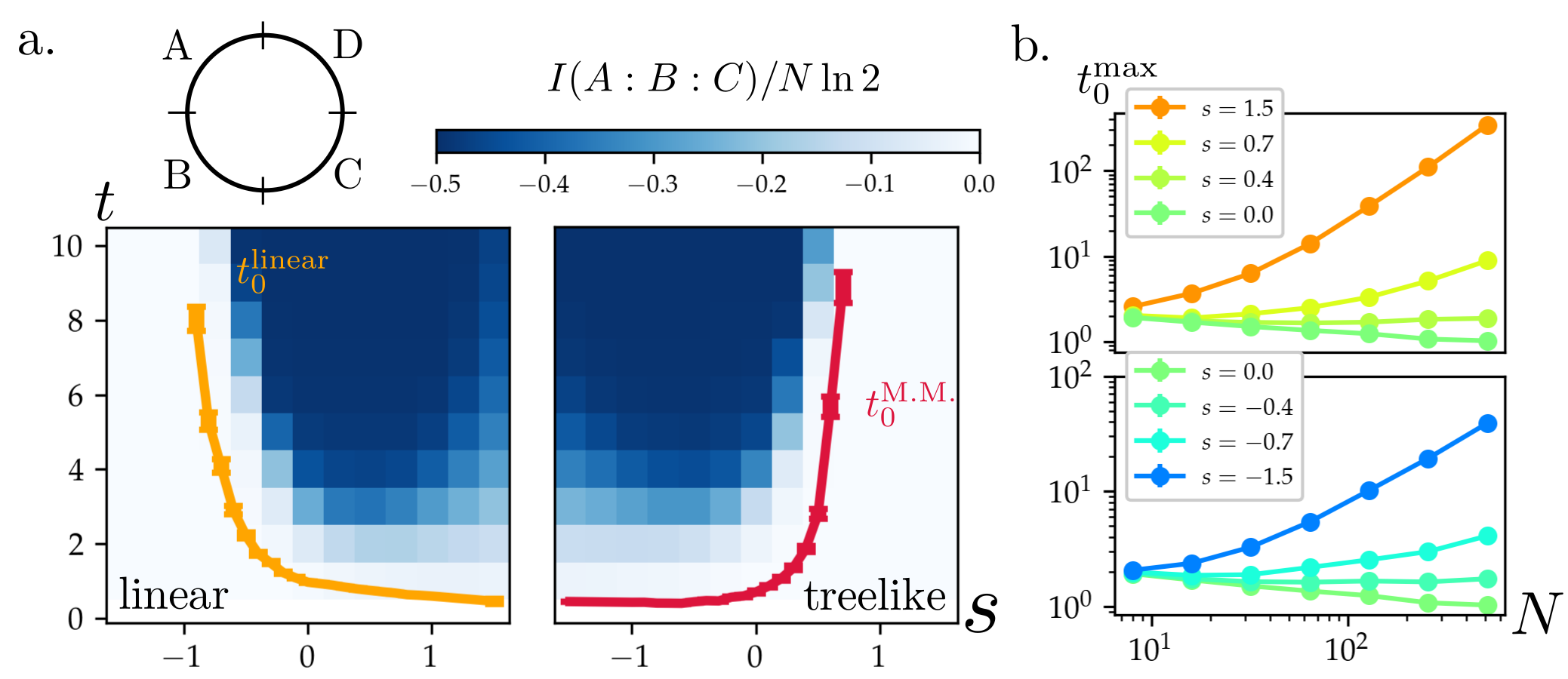} %[width=1.0\columnwidth]
   \caption{Characterizing Scrambling using Tripartite Mutual Information.
   \label{fig:tripartite}
      (\textbf{a})~The tripartite mutual information $I(A:B:C)$ between three contiguous regions of size $N/4$ (top left)
      of the output qubits for $s=-1.5, \ldots, 1.5$ in steps of $\Delta s = 0.25$, in linear (left) and treelike (right) geometries for $N = 512$.
      An initial plateau $I(A:B:C)=0$ (white) persists for long times when the interactions are local in a given geometry.
      At long times, information eventually scrambles across the entire system, leading to substantial negativity (dark blue). 
      We estimate the timescale of the initial plateau by finding the time $t_0$ required for the system to reach one bit of negativity $I(A:B:C) = -\ln 2$ when the regions $A,B,C$ are chosen contiguously from either the linear (orange) or treelike (red) geometry.
      (\textbf{b}) The timescale $t_0$ for $s = \pm 0.0, 0.4, 0.7, 1.5$ grows rapidly with system size $N$ in both of the local regimes 
      $\magn{s} > 1$, but is insensitive to system size near the fast-scrambling limit $s = 0$.
      Error bars are shown or are smaller than data points. Results are averaged over $10^{3}$ circuit realizations with 
      product state $\ket{0}^{\otimes N}$ as the initial state (\appref{app:TDTMI}).}
    %   The timescale required to reach one bit of negativity $I(A:B:C) = -\ln 2$ varies dramatically with the exponent $s$
    %   (b)~Crossing of $I(A:B:C)$ for different system sizes ($N=2^{5}\cdots2^{9}$,light to dark) 
    %   at $s=-0.5$ (top) and $s=0.2$ (bottom) in linear (blue) and treelike (purple) geometries.
    %   When the interactions are no longer local in a particular geometry, there exists a point in time which they all cross.
    %   (c)~The maximum duration of initial plateau at $I(A:B:C)=0$ between the linear and treelike geometries
    %   ($t^{\max}_0=\max\{t^{\mathrm{Arch.}}_0,t^{\mathrm{non-Arch.}}_0\}$) 
    %   for $s=\mp0,0.2,0.4,0.5,0.6,0.7,1,2$ (light to dark) with respect to the system sizes $N=2^{3},\cdots,2^{9}$.
    %   The conversion of $t^{\max}_0$ near $s=0$ indicates that there exists a weak-scrambling time $t_{c,\mathrm{weak}}$
    %   which is system-size invariant for suffices only large system size.
    %   Data points are obtained by averaging of up to $10^{3}$ trajectories. 
    %   Error bars in some points are smaller than the markers.
    %   The symbol $\mp$ implies negative values of $s$ ($s\leq 0$) are on the top panel 
    %   and positive values of $s$ ($0\leq s$) are on the bottom panel. 
\end{figure}

In our sparse Clifford circuits, we find that the time dependence of the tripartite mutual information varies dramatically depending on the value of $s$ as shown in Figure \ref{fig:tripartite}a.
For large $|s| > 1$, the locality of the interactions in both the linear and treelike geometries results in an effective lightcone that prevents system-wide scrambling at short times, which appears in Figure \ref{fig:tripartite}a as a long plateau near $I(A:B:C) = 0$. Only after timescales extensive in $N$, when the local lightcone has spread to cover the entire system, do we observe appreciable negativity in the tripartite mutual information.
Near $|s|=0$, by contrast, the tripartite mutual information becomes negative in a timescale of order  $t_* \sim \mathcal{O}(1)$
%\sout{consistent with the fast-scrambling conjecture \cite{sekino2008,lashkari2011towards,bentsen2019fast}.}}
due to the emerging hierarchical structure of the circuit (see \appref{app:t0scaling}).
% The local lightcones prevent entanglement from growing faster than the nonlocally across the three disjoint subsystems. 
At long times the tripartite mutual information approaches the expected steady-state value of $I(A:B:C)\approx -\frac{N}{2}\ln 2$ 
(\appref{app:ScramblingLimit}).

\textls[-31]{We can quantitatively extract the timescale required for system-wide entanglement to build up in our circuits 
by computing the time $t_0$ required to reach one bit of negativity in the tripartite mutual information 
in a given geometry.
With qubits arranged in a conventional linear geometry (Figure \ref{fig:tripartite}a left, orange),
this timescale grows as $t^{\mathrm{linear}}_0 \sim N$ for  $s < -1$ for large system sizes. 
In the treelike geometry, for $1 < s$ the time $t_0^{\mathrm{treelike}}$ 
grows like $t^{\max}_0\sim N^{\beta}$ with $\beta \sim s+0.1$ (Figure \ref{fig:tripartite}a right, red).
Finally, at the crossover point $s=0$ we find that $t_{0}^{\max}=\max\{t_0^{\mathrm{linear}},t_0^{\mathrm{treelike}}\}$
converges towards $t=1$ as shown in Figure \ref{fig:tripartite}b (see \appref{app:t0scaling} for further details and discussion). 
This quick loss of locality of the interactions in the two mathematically incompatible geometries near $s=0$ 
further confirms the fast-scrambling nature of this circuit.} 

\section{Characterizing the Many-Body Lightcone via Teleportation}
\label{sec:telelightcone}

% \sectabst{Greg: Map out the scrambling lightcone by computing mutual information $I(i:j)$ between input qubit $i$ and output qubit $j$ as in arXiv:2110.06963.}

One can obtain a striking picture of the system's local geometry by studying its lightcone, which governs the propagation of quantum information in general quantum systems \cite{lieb1972finite,hastings2010locality,bentsen2019treelike,tranLiebRobinsonLightCone2021}. In 1 + 1d systems with local interactions this lightcone is linear,
such that information can reliably propagate between spatially separated sites $i,j$ only if they lie inside the lightcone $\magn{i-j} < v_s t$ for some finite velocity $v_s$. The velocity $v_s$ is sometimes called the Lieb--Robinson velocity or the butterfly velocity depending on context, and its numerical value depends on the microscopic details of the system.
Previous work has studied lightcones in tunable sparse models using free-fermion and MPS numerics \cite{bentsen2019treelike} as well as via field-theory methods \cite{gubser2018continuum,gubser2019mixed}. In particular, these previous results indicated the emergence of a linear lightcone $\magn{i-j} < v_s t$ for sufficiently negative $s < 0$ and a treelike lightcone $\magn{i-j}_2 < v_s t$ characterized by the treelike 2-adic distance $\magn{\cdot}_2$ for sufficiently positive $s > 0$. Here we present evidence for similar emergent lightcones in our family of sparse Clifford circuits, which allows us to probe the linear-treelike transition in the strongly-interacting limit.

\begin{figure}[H]
    \includegraphics[width=0.92\textwidth]{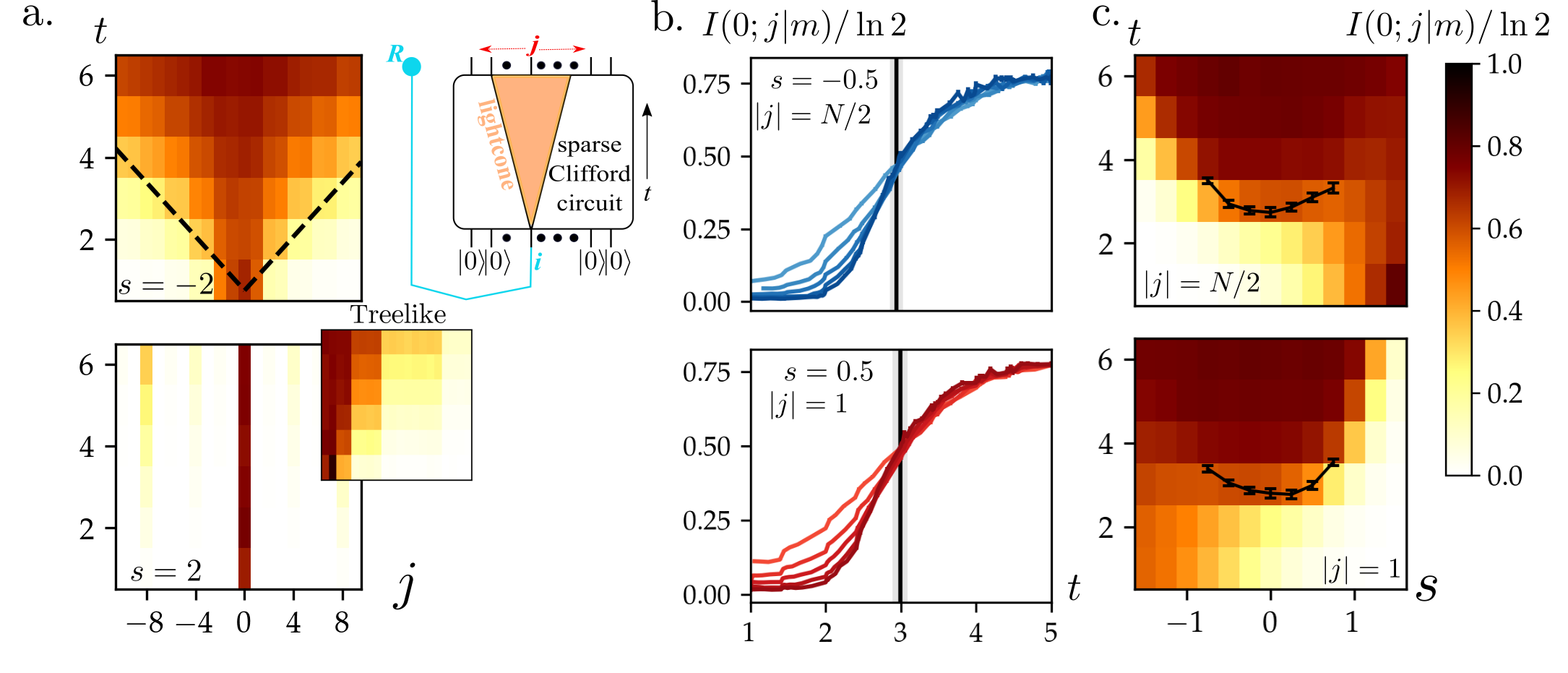} %[width=1.0\columnwidth]
    \caption{\textls[-15]{Characterizing tunable geometry via teleportation. 
    We probe the many-body lightcone in sparse Clifford circuits by maximally entangling a reference qubit with a fixed input qubit $i$ and computing the mutual information $I(i;j \rvert m)$ between the reference and a particular output qubit $j$ as a function of distance $\magn{i-j}$ (\textbf{a}). The mutual information characterizes the fidelity of teleporting a qubit from $i$ to $j$, which is possible only after the input qubit $i$ has been sufficiently scrambled into qubits lying inside the lightcone. When $s = - 1$ we find a linear lightcone (top), where strong scrambling inside the lightcone (black dashed, $v_s \approx 3.0$) enables successful teleportation of quantum information from $i$ to $j$. When $s = 1$ this lightcone is apparently destroyed (bottom), but a rearrangement of the sites by the Monna map $\mathcal{M}$ yields an emergent treelike lightcone as measured by the 2-adic distance $\magn{i-j}_2$ (inset). For $-1 < s < 1$ we find a finite-time phase transition (\textbf{c}) in the teleportation fidelity. To probe this phase transition for $-1 < s \leq 0$ (top) we consider teleportation between qubits maximally separated in linear (Euclidean) distance $\magn{i-j} = N/2$; for $0 \leq s < +1$ (bottom) we consider qubits maximally separated in treelike 2-adic distance $\magn{i-j}_2 = 1$. The critical points $t_c(s)$ of this transition are extracted at each $s$ via finite-size scaling analysis (\textbf{b}) as explained in the main text. Error bars are shown in black.}}
    \label{fig:telelightcone}
\end{figure}

To probe the many-body lightcone, we ask whether it is possible to teleport a single qubit of information from a particular location $i$ at the input of the circuit to another location $j$ at the output \cite{bao2021finite} as illustrated in Figure \ref{fig:telelightcone}a. Teleportation via the standard Hayden--Preskill--Yoshida--Kitaev mechanism succeeds when the input is strongly scrambled into an output region containing the output qubit $j$ \cite{hayden2007black,yoshida2017efficient,Yoshida2019Disentangling}. We therefore expect teleportation to succeed inside the lightcone $\magn{i-j} < v_s t$ and to fail outside of it.
To test this in our tunable Clifford circuits, we maximally entangle a single reference qubit $R$ with the input qubit $i$ as illustrated in Figure \ref{fig:telelightcone}a. We then measure the mutual information
\begin{equation}
    I(i;j \rvert m) = \left(S_i + S_j - S_{ij} \right) \rvert_m
\end{equation}
between the input and output qubit $j$, conditioned on performing projective measurements in the Pauli-$Z$ basis on all qubits except qubits $R,j$ \cite{bao2021finite}.
% Inside the many-body lightcone $\magn{i-j} < vs. t$, information is maximally scrambled among all $k$ qubits inside the lightcone, and we find nonzero mutual information $I(i:j) > 0$, signalling successful teleportation. Outside the many-body lightcone, there are no correlations between the reference qubit and the output qubit, and so teleportation fails.

\textls[-15]{When $s < 0$ this procedure yields a linear lightcone at short times, as shown in \mbox{Figure \ref{fig:telelightcone}a} (black dashed), with a velocity $v_s \approx 3.0$. By contrast, when $s > 0$ this lightcone is apparently destroyed, and information can teleport between distantly-separated sites $i,j$ only when they are separated by large powers of 2. Upon rearranging the sites by the Monna map, however, we recover a treelike lightcone as measured by the treelike 2-adic distance $\magn{i-j}_2 < v_s t$ (inset). These emergent lightcones demonstrate that our circuits behave locally for large $\magn{s}$, but that the notion of locality depends strongly on the sign of $s$, yielding local dynamics in either conventional linear geometry when $s < 0$ or treelike geometry when $s > 0$.}

\textls[-5]{Near the crossover point $s = 0$ between these two geometries, we expect both local lightcones to collapse, yielding fast scrambling dynamics where information rapidly spreads throughout the system \cite{bentsen2019treelike}. These fast-scrambling dynamics should exhibit a finite-time teleportation phase transition similar to that found recently in all-to-all random circuits~\cite{bao2021finite}. We study this phase transition as a function of $-1 < s < 1$ in Figure \ref{fig:telelightcone}c, using pairs of qubits $i,j$ that are either maximally separated in linear distance $\magn{i-j} = N/2$ (top) or maximally separated in treelike distance $\magn{i-j}_2 = 1$  (bottom).
Both choices yield a clear phase transition at a critical time $t_c(s)$ for circuits near $s = 0$. The critical time $t_c(s)$ is extracted at each value of $s$ by determining the crossing point of the teleportation fidelity at early times in a finite-size scaling analysis (Figure \ref{fig:telelightcone}b).
% In this regime, we extract the critical time $t_c(s)$ by performing finite-size scaling analysis to find the crossing point of the mutual information $I(i;j|m)$ as we vary the system size $N$ (\figref{fig:telelightcone}b).
These critical points are overlaid in black on Figure \ref{fig:telelightcone}c. By contrast, similar finite-size scaling analysis in the local regimes $1 \leq |s|$ yield no such crossings, indicating that there is no longer a finite-time phase transition in these regimes.
Together, these results provide clear evidence of emergent lightcones in the linear $s < 0$ and treelike $s > 0$ regimes, separated by the nonlocal regime characterized by a finite-teleportation-time facilitated by the fast scrambling dynamics near $s=0$.}

\section{Conclusions and Outlook}
\label{sec:discussion}

% \sectabst{Greg: Wrap things up, give a summary, some thoughts for future work.}

\textls[-15]{In this paper, we studied tunable effective geometries that emerge in sparse circuits, using Clifford circuits as toy models. To characterize the effective geometries in these circuits we numerically computed Renyi entropies of contiguous regions of output qubits $A$ as a function of time, where the term `contiguous' was defined either according to the linear (Euclidean) metric $\magn{i-j}$ or according to the treelike 2-adic metric $\magn{i-j}_2$. In Section \ref{sec:arealawvolumelaw}, we found area-law entanglement at short times $t < N$ in both of the local regimes $\magn{s} > 1$, indicating linear entanglement growth for negative $s < 0$ and treelike entanglement growth in a treelike geometry for positive $s > 0$. 
We confirmed these findings in Section \ref{sec:trimutualinfo} where we studied the timescale for quantum information to become scrambled as quantified by the negativity of the tripartite mutual information $I(A:B:C)$. There we found that either a linear ($s < 0$) or treelike ($s > 0$) lightcone prevents system-wide scrambling earlier than a time $t_0 \sim \mathcal{O}(N^a)$ in the local regimes $\magn{s} > 1$. Finally, in Section \ref{sec:telelightcone} we explicitly mapped the system's many-body lightcone by considering the fidelity $I(i;j|m)$ of teleporting a single qubit of information from an input qubit $i$ to an output qubit $j$. Here we found that teleportation succeeds if the ouput qubit $j$ resides inside the many-body lightcone of the intput qubit $i$, and fails if $j$ lies outside this lightcone. We demonstrated that this lightcone has a linear structure for $s < 0$ and a treelike structure for $s > 0$. Near the crossover point $s = 0$ both of these lightcones collapse and teleportation succeeds in a finite time $t_c(s)$ that depends on the exponent $s$.}

While in this work we were able to extract the teleportation time $t_c(s)$ for $-1 < s < 1$ in finite-size scaling analysis, future work may explore how this phase disintegrates near $\magn{s} = 1$, where the teleportation time diverges in the thermodynamic limit. Using the teleportation time $t_c(s)$ as a diagnostic, one might expect to observe a phase transition from a fast scrambling phase to a slow scrambling phase as a function of the parameter $s$ near the points $\magn{s} = 1$. It would also be interesting to study similar questions in Brownian qubit or fermion models, where analytical results are often possible \cite{zhang2021emergent,bentsen2021measurement}. In particular, these Brownian models typically lead to a description of entanglement dynamics in terms of effective field theories. Defining one of these theories on a sparse coupling graph such as the one considered here might allow us to make contact between the entanglement dynamics studied here and analytical field-theory approaches \cite{gubser2018continuum}.
\vspace{6pt}

% for linearly contiguous regions $A$ when $s < 0$, indicating an underlying linear geometry. Similarly, we found area-law entanglement for treelike contiguous regions when $s > 0$, indicating a treelike geometry in this limit. by contrast, all other choices of region $A$ exhibit volume-law scaling.

%%%%%%%%%%%%%%%%%%%%%%%%%%%%%%%%%%%%%%%%%%
\authorcontributions{Numerical simulations, T.H. and S.K.; Conceptualization and writing, G.B., T.H., S.K. and A.J.D. All authors have read and agreed to the published version of the manuscript.}

\funding{Work at the University of Strathclyde was supported by the EPSRC Programme Grant DesOEQ (EP/P009565/1), the EPSRC Quantum Technologies Hub for Quantum Computing and simulation (EP/T001062/1), the European Union’s Horizon 2020 research and innovation program under grant agreement No. 817482 PASQuanS, and AFOSR grant number FA9550-18-1-0064. G.B. is supported by the DOE GeoFlow program (DE-SC0019380). }

\institutionalreview{Not applicable.}

\informedconsent{Not applicable.}

\dataavailability{Data for this manuscript can be found in {open-source at} \url{https://doi.org/10.15129/9abc9cc4-1a5a-4725-8b92-1c71f3033d17} (accessed on 22 February 2022).
} 

\acknowledgments{Results were obtained using the ARCHIE-WeSt High Performance Computer (\url{www.archie-west.ac.uk}) (accessed on 22 February 2022)
based at the University of Strathclyde.}

\conflictsofinterest{The authors declare no conflict of interest.} 

%}}}

%{{{ Appendix 

%%%%%%%%%%%%%%%%%%%%%%%%%%%%%%%%%%%%%%%%%%
%% Optional
\appendixtitles{yes} % Leave argument "no" if all appendix headings stay EMPTY (then no dot is printed after "Appendix A"). If the appendix sections contain a heading then change the argument to "yes".
\appendixstart
\appendix

\section[\appendixname~\thesection]{Stabilizer Formalism and Clifford Circuits}
\label{app:stabilizers}

% \sectabst{Greg: Explain stabilizers and Cliffords}

Here we review the stabilizer formalism and Clifford circuits, powerful tools for understanding a restricted class of many-body quantum systems where the dynamics can be efficiently simulated on a classical computer \cite{gottesman1998heisenberg,aaronson2004improved}. Consider the Pauli group $\mathcal{P}(Q)$ of all Pauli strings $g = \prod_i X_i^{x_i} Z_i^{z_i}$ acting on a system $Q$ of qubits labeled $i = 0,1,\ldots,N-1$, where the bits $x_i,z_i = 0,1$ specify whether a particular Pauli operator $X_i,Z_i$ is present in the string. We define an Abelian \emph{stabilizer} subgroup $\mathcal{S} \leq \mathcal{P}(Q)$ which is generated by a set $\mathcal{G} = \{g_1,g_2,\ldots,g_m\}$ of linearly-independent, mutually-commuting Pauli strings $[g_{\ell},g_{\ell'}] = 0 \ \forall \ell,\ell' = 1,\ldots,m$. A stabilizer group with $m \leq N$ independent generators has order $\magn{\mathcal{S}} = 2^m$. Given a stabilizer group $\mathcal{S}$ one can construct a \emph{stabilizer state}
\begin{equation}
    \rho_Q(\stabs) = 2^{-\magn{Q}} \sum_{g \in \mathcal{S}} g
    \label{eq:codestate}
\end{equation}
which is the unique density matrix \emph{stabilized} by $\mathcal{S}$: $g \rho_Q(\mathcal{S}) g = +\rho_Q(\mathcal{S})$ for all elements $g \in \mathcal{S}$. If we specify a `complete' set of $m = N$ stabilizers, then \eqref{eq:codestate} is a rank-1 projector onto the unique pure state $\ket{\psi}$ stabilized by the group $\mathcal{S}$ (i.e., $g\ket{\psi} = +\ket{\psi}$ for all $g \in \mathcal{S}$).

\subsection{Classical Simulation of Clifford Circuits}

The Clifford group $\mathcal{C}$ on $N$ qubits is the group of all unitary operators which transform Pauli strings into other Pauli strings (in other words, elements of $\mathcal{C}$ map 
$\mathcal{P}(Q)$ onto itself).
Therefore elements of the Clifford group also map every stabilizer state $\rho_Q(\mathcal{S})$ onto another stabilizer state $\rho_Q(\mathcal{S'})$ with stabilizer group $\mathcal{S'}$. Gottesman and Knill demonstrated that one can classically simulate the evolution of these stabilizer states under the action of the Clifford group by mapping the dynamics onto linear algebra operations over the field $\mathrm{GF}(2)$ (i.e., binary numbers mod 2) \cite{gottesman1998heisenberg,aaronson2004improved}.
% This transformation is equivalent to the transformation of state $\ket{\Sigma}$ to $\ket{\Sigma'}$.
% Therefore, the evolution of a stabilizer state under the unitaries from the Clifford group can be 
% tracked by keeping track of how the initial stabilizer transforms.
% The evolution with unitaries from the Clifford group 
% is proven to be able to be computed classically in polynomial time 
% by Gottesman and Knill.
This is accomplished by mapping the stabilizer group $\mathcal{S}$ onto a binary $m \times 2N$ matrix $\mathcal{M}$, where each row $\ell = 1,\ldots,m$ of the matrix is a binary string $\vec{b}_{\ell} = (x_1,\ldots,x_N,z_1,\ldots,z_N)$ corresponding to the generator $g_{\ell} = \prod_i X_i^{x_i} Z_i^{z_i}$.

\textls[-12]{Equipped with a binary matrix $\mathcal{M}$ describing the state of our system at any fixed time, we now show that time-evolution under elements of the Clifford group corresponds to performing simple row and column linear algebra operations on this matrix. The Clifford group is generated by the Hadamard ($H$), Phase ($P$), 
and controlled-NOT (CNOT) gates, so we proceed by describing the action of each of these elementary gates  \cite{gottesman1998heisenberg,aaronson2004improved,Selinger2015}.
The Hadamard gate $H_i$ on site $i$ exchanges the operators $Z_i \leftrightarrow X_i$ which corresponds to swapping columns $i$ and $i+N$ in the matrix $\mathcal{M}$.
The Phase gate $P_i$ on site $i$ exchanges the operators $X_i \leftrightarrow Y_i$; in the matrix $\mathcal{M}$ this is equivalent to setting column $i+N$ equal to the sum (mod 2) of columns $i$ and $i+N$.
Finally, a controlled-NOT gate CNOT$_{ij}$ applied between a control qubit $i$ and a target qubit $j$ is equivalent to transforming Pauli strings according to the following four rules:
$X_i I_j \rightarrow X_i X_j$, $I_i X_j \rightarrow I_i X_j$, $Z_i I_j \rightarrow Z_i I_j$, and $I_i Z_j \rightarrow Z_i Z_j$, which correspond to setting column $j$ equal to the sum of columns $i$ and $j$ (mod 2), and setting column $i+N$ equal to the sum of columns $i+N$ and $j+N$ (mod 2).
With these $H_i$, $P_i$, and CNOT$_{ij}$ gates in hand, we may systematically generate all $N$-qubit operators in 
$\mathcal{C}$ using standard algorithms \cite{Selinger2015}.}

\subsection{Reduced Density Matrices and Entanglement Entropy}
Given a stabilizer state \eqref{eq:codestate} and its associated stabilizer matrix $\mathcal{M}$, we may readily compute entanglement entropies by performing some simple linear algebra. The reduced density matrix $\rho_A(\stabs)$ for a subregion $A \subset Q$ is obtained by tracing out $\bar{A}$, which is equivalent to tracing out the Pauli operators which belong to $\bar{A}$. This yields

\begin{align}
    \rho_{A}(\stabs_A) = \mathrm{Tr}_{\bar{A}} [ \rho_Q(\stabs) ] 
    = 2^{\magn{\bar{A}} - \magn{Q}}\sum_{g\in\mathcal{S}} \mathrm{Tr}_{\bar{A}} [g] = 2^{-\magn{A}}\sum_{g_A\in\mathcal{S}_A} g_{A},
\end{align}
where $\mathcal{S}_A\subset\mathcal{S}$ is the subgroup of stabilizers $g_A$ which have no support on $\overline{A}$
\cite{Nahum2017}. Given a stabilizer matrix $\mathcal{M}$, tracing out the subregion $\overline{A}$ is equivalent to simply discarding the columns corresponding to Pauli operators $X_j,Z_j$ for $j \in \overline{A}$, and then performing row-reduction on the remaining columns to find a linearly independent set of generators~$g_A$.

The Renyi entropy of any stabilizer state $\rho_A(\stabs_A)$ is determined simply by the binary rank ($\mathrm{rank}_{\gft}$) of its associated matrix $\mathcal{M}$. To see this, let us calculate the purity
\begin{align}
    \tr{\rho_A^2(\stabs_A)} = 2^{- 2 \magn{A}} \sum_{g,g' \in \stabs_A} \tr{g g'} = 2^{- \magn{A}} \cdot \magn{S_A} 
    = 2^{\mathrm{rank}_{\gft}(\mathcal{M}) - \magn{A}}
\end{align}
where $\magn{S_A} = 2^{\mathrm{rank}_{\gft}(\mathcal{M})}$ is the cardinality of the stabilizer group and we have used the fact that $\tr{g^2} = 1$, while $\tr{g g'} = 0$ for any two nonequal Pauli strings $g \neq g'$. Therefore the Renyi entropy is simply
\begin{equation}
    S_A = -\ln \tr{\rho_A^2(\stabs)} = \ln 2 \left(\magn{A} - \mathrm{rank}_{\gft}(\mathcal{M}) \right).
\end{equation}

\section[\appendixname~\thesection]{Normalization Factor $\mathbf{\emph{J}}_\mathbf{\emph{s}}$ of the Sparse Probability Distribution}
\label{app:normalization}
There are $2\log_2 N-1$ unique distances which the gates can be applied to a site in the sparse Clifford circuit.
They are $\log_2 N-1$ pairs of distance $r=2^{0,1,\cdots,\log_2 N-2}$ gates and one $r=N/2$ gate.
Unlike other stochastic models with random gate applications \cite{sekino2008,nahum2021measurement},
our circuit is deterministic up to the allowed gate distance at a particular layer of a circuit with a period of 
$2\log_2 N-1$ layers.

In order to compare the circuit with different number of qubits $N$ and exponent $s$, 
we fix the number of gates applied to the qubits per site per period of the circuit. 
Throughout this paper, we set this number to be $1$. 
This naturally gives us the normalization of the sparse probability distribution, from which we choose the gate distances. It is
\begin{align}
   \frac{1}{J_s} = \mathcal{N}_s = \frac{1}{2}\left( \left(\frac{N}{2}\right)^s+2\sum_{k=0}^{\log_2 N-2} 2^{ks} \right).
\end{align}

\textls[-9]{The probability of applying a gate of distance $r$ is, hence, governed by the exponent $s$ as}
\begin{align}
   p(r,s)=J_sr^s = \frac{r^s}{\mathcal{N}_s}.
\end{align}

\section[\appendixname~\thesection]{$\mathbf{\emph{p}}$-Adic Numbers and the Monna Map}
\label{app:monnamap}

\textls[-15]{The real numbers $\mathbb{R}$ are defined as a natural extension of the rational numbers, but this is not the only possible choice. The $p$-adic numbers $\mathbb{Q}_p$ are an alternative extension of the rational numbers, each associated with a prime number $p$ \cite{FQ_1997_padic}. 
Whereas distances in the reals $\mathbb{R}$ are measured using the usual Euclidean norm $|\cdot |$, 
distances between $p$-adic numbers are measured by the \emph{p}-adic norm $|\cdot |_p$.
The idea behind the $p$-adic norm is to use the multiplicity of the prime $p$ in the factorization of numbers. The definition of the $p$-adic metric is}
\begin{equation}\label{padic_norm}
    \magn{i-j}_{p}=\begin{cases} 
      p^{-v_p(i-j)} & (i-j \neq 0)\\
      0 & (i-j = 0),
   \end{cases}
\end{equation}
\textls[-15]{where $v_p$ is the number of times the prime factor $p$ is present in the factorization of the irreducible fraction $i-j$. Perhaps surprisingly, this definition satisfies the usual axioms required of a metric, and therefore serves as a useful notion of distance.
One of the notable and crucial properties of the $p$-adic numbers is their ultrametricity as mentioned in \mbox{Section \ref{sec:arealawvolumelaw}}.
Obeying this strong triangle inequality naturally gives rise to a hierarchical structure, 
where decreasing $v_p$ implies moving deeper in to the tree in Figure \ref{fig:overview}c.
This infinite family of $p$-adic numbers has found its way  into field theories, starting with Dyson's one-dimensional model~\cite{dysonExistencePhasetransitionOnedimensional1969} and the ensuing studies in \cite{Bleher_1973,Lerner_1989}. In our analysis, the emergent treelike geometry in the $s>0$ limit can be understood by using the $(p=2)$-adic metric as a tool
for capturing the treelike structure in our circuit following the analysis on the closely related models \cite{gubser2018continuum,bentsen2019treelike}. }

%\tomohiro{Redundant perhaps?} One can define the 2-adic norm from \eqref{padic_norm} using $p=2$. For example, $\magn{4}_{2}=\magn{2}^{-2}=1/4$. This means that, if we have $N=8$, the sites $0$ and $4$ are hierarchically closer to each other than $0$ and $1$ which is counter-intuitive to what we see in linear (Euclidean) geometry.   

\subsection*{Monna Map}
To arrange integers sequentially in using the $2$-adic norm, 
one can use a map $\mathcal{M}$ called Monna map introduced by Monna \cite{MONNA1952}.
%Here, we employ a discrete version of Monna map for understanding sparse coupling patterns \cite{gubser2018continuum} 
%which is in contrast to the original continuous map established by Monna \cite{MONNA1952}. 
Let
\begin{equation} \label{defn_L}
    N=2^{n}
\end{equation}
where $n$ is an integer, 
then $x \in \{0,1,2,\ldots,N-1\}$ can be expressed as
\begin{equation}\label{defn_x}
    x = \sum_{q=0}^{n-1}\bar{x}_{q}p^{q},
\end{equation}
where $\bar{x}_{q} \in \{0,1,\ldots,p-1\}$.
Then the Monna map $\mathcal{M}$ is defined as
\begin{equation}\label{discrete_monna}
    \mathcal{M}(x) \equiv \sum_{q=0}^{n-1} \bar{x}_{n-1-q}p^{q},
\end{equation}
which means that we reverse the digits in the base $p$ of expansion $x$. Using this definition of Monna Map for $p=2$, we reverse the binary equivalent of the argument. This leads to the treelike ordering of qubits as illustrated in Figure \ref{fig:overview}c.

\section[\appendixname~\thesection]{Curve Fits for the Timescale $\mathbf{\emph{t}}_{\mathrm{\mathbf{\emph{vol.}}}} $ }
\label{app:Linear_Fits}

\begin{figure}[H]
   \includegraphics[width=0.84\textwidth]{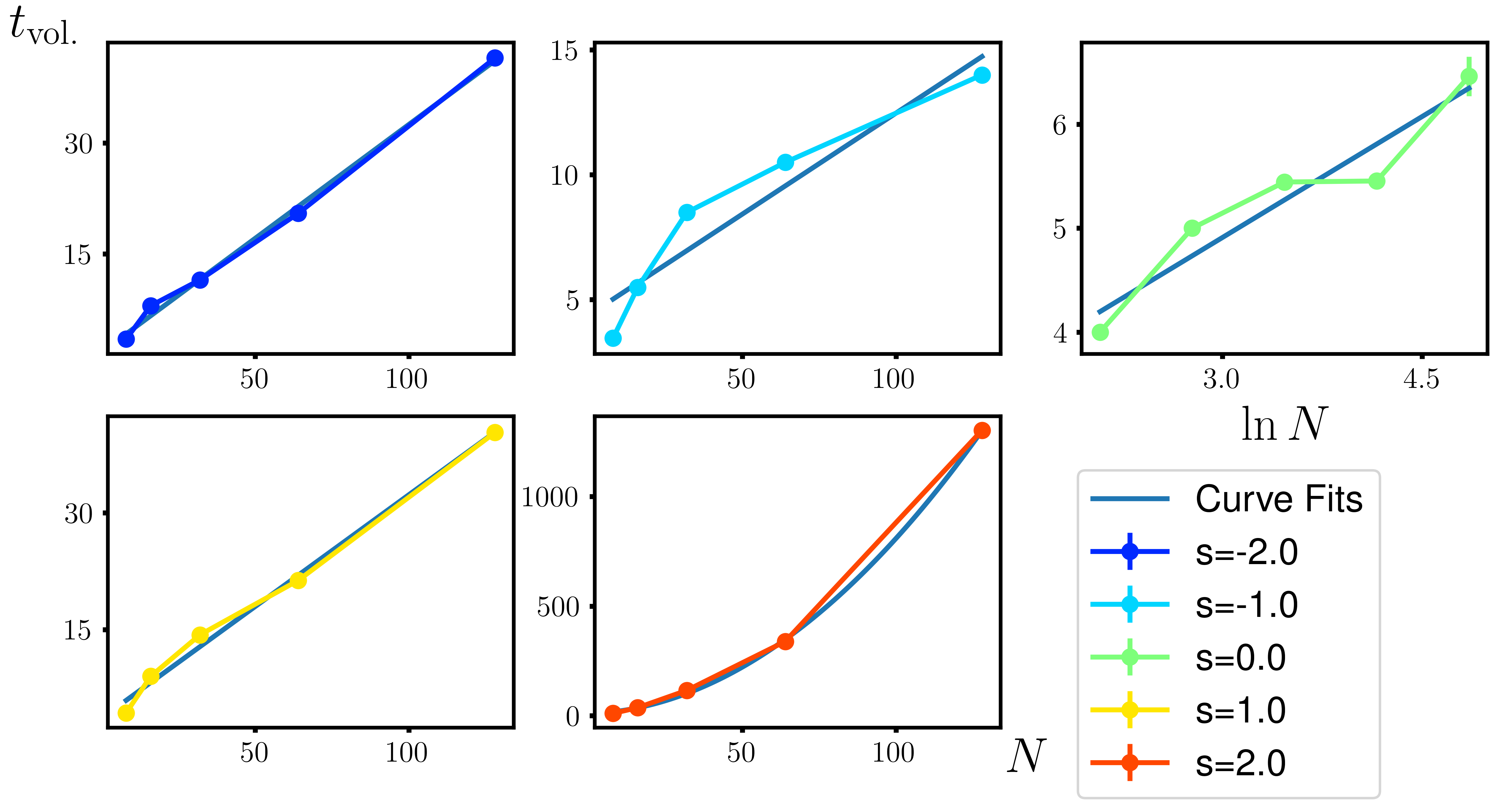}
   \caption{
      Curve fits for timescale $t_{\mathrm{vol.}} $. To strengthen our observation of different growth rates of timescale $t_{\mathrm{vol.}}$ with system size $N$ and tunable parameter $s$, we use linear fits on $t_{\mathrm{vol.}}$ and $N$ data for $s\leq -1$, and $t_{\mathrm{vol.}}$ and $\ln{N}$ data for $s=0$. In the treelike $s>0$ limits, we use the $s{\mathrm{th}}$ degree polynomial to show that the timescale $t_{\mathrm{vol.}}$ grows polynomially in the system size.
   \label{fig:LinearFits}}
\end{figure}

\textls[-15]{In Section \secref{sec:arealawvolumelaw}, we claimed that the timescale $t_{\mathrm{vol.}} $ required to reach the volume law entanglement grows linearly with system size $N$ when $s\leq -1$, and polynomially with system size when $s>-1$, except at $s=0$, where $t_{\mathrm{vol.}}  \sim \ln{N}$. In Figure \ref{fig:LinearFits}, we show the curve fits for the data, hence verifying the different scaling of time $t_{\mathrm{vol.}}$ with system size for different values of $s$. When $s\leq -1$, we linear fit $t_{\mathrm{vol.}}$ and $N$, in the treelike regime ($s>0$), we fit the $s{\mathrm{th}}$ degree polynomial in \emph{N} and at the crossover point $s=0$, we linear fit $t_{\mathrm{vol.}}$ and $\ln{N}$.}
\vspace{-6pt}

\section[\appendixname~\thesection]{Time Dependence of Tripartite Mutual Information}
\label{app:TDTMI}

\begin{figure}[H]
   \includegraphics[width=\textwidth]{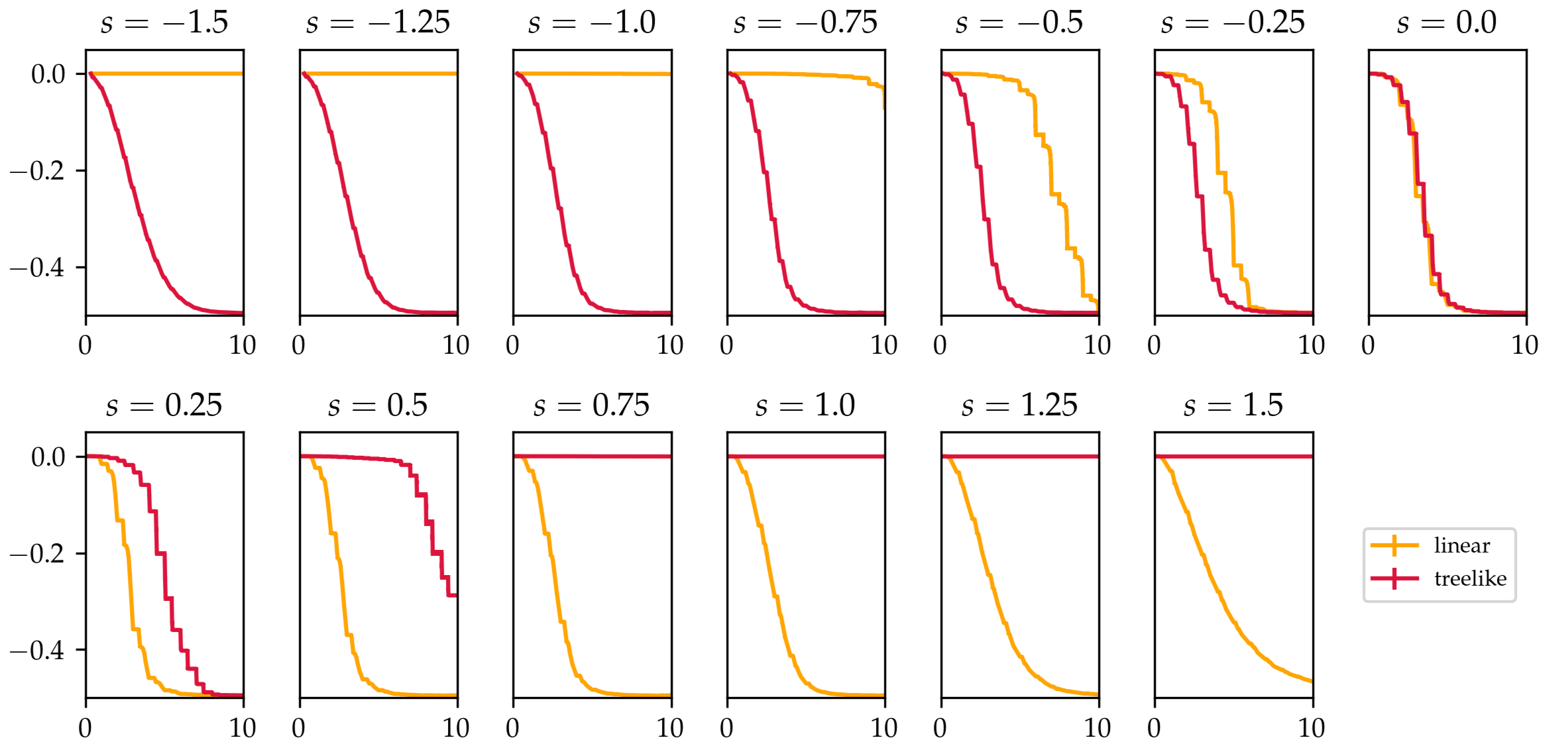}
   \caption{
      Time dependence of tripartite mutual information.
   The time dependence of tripartite mutual information $I(A:B:C)/N \ln 2$
   evolved with sparse Clifford circuit up to $t=10$ with system size $N=512$ 
   for $s=-1.5,-1.25,-1.0,-0.5,-0.75,-0.25,0,0.25,0.5,0.75,1.0,1.5$.
   It is calculated with the four contiguous systems of size $L_{\mathrm{sub}}=N/4$
   for linear and treelike geometries (orange and red, respectively). 
   The mean values are calculated and the errors are estimated from $1\times10^{2}$~trajectories. 
   \label{fig:ATPIerror}}
\end{figure}

The color plot in Figure \ref{fig:tripartite}a is a plot of tripartite mutual information of a state evolved in 
the sparse Clifford circuit with system size $N=512$.
It is plotted for the values of $s$: $s=-1.5,-1.25,-1.0,-0.5,-0.75,-0.25,0,0.25,0.5,0.75,1.0$, and $1.5$.
In Figure \ref{fig:ATPIerror}, we show the time dependence of the quantity 
in linear and treelike geometries (orange and red, respectively) from $t=0$ to $t=10$.
The mean values are obtained and errors are estimated from $1\times10^{2}$ trajectories.

\section[\appendixname~\thesection]{The Limiting Behavior of $\mathbf{\emph{t}}_\mathbf{0}^{\mathrm{\mathbf{linear}}}$ at $\mathbf{\emph{s}}=\mathbf{0}$ for Large System Sizes}
\label{app:t0scaling}
In this section, we show that the at $s=0$, 
the tripartite mutual information $I(A:B:C)$ in contiguous subsystems of size $N/4$ of a system of $N$ sites in the linear geometry
as discussed in Section \secref{sec:trimutualinfo},
becomes negative in the time scale of order $1$ (i.e., $t^{\max}_0 \sim \mathcal{O}(1)$) in the limit of large system size. 

\begin{figure}[H]
   \includegraphics{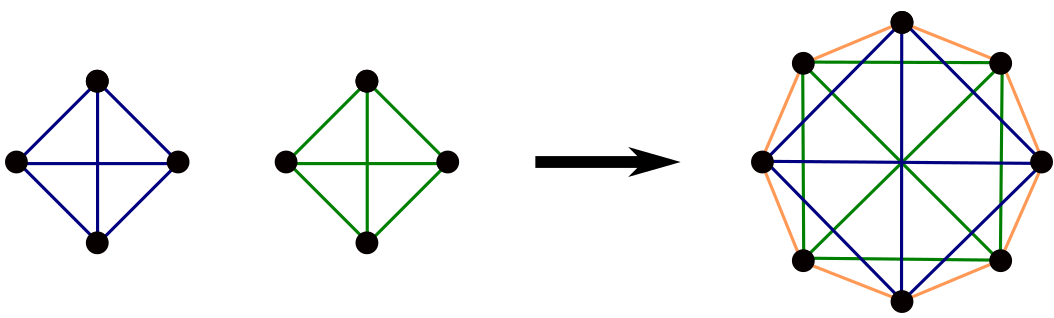}
   \caption{Hierarchical Construction of the underlying coupling graph of the sparse Clifford circuit.
      The coupling graph, which the circuit obeys which in comes to the possible position of the gates,
      is governed by the sparse coupling graph. The sparse coupling graph of size $N$ (right, $N=8$) can be 
      constructed from two coupling graphs of size $N/2$ (left, $N/2=8$) by interleaving the sites and connecting them with the 
      nearest neighbor coupling (right, orange).
   \label{fig:hierarchy}}
\end{figure}

The sparse Clifford circuit of size $N$ can be constructed hierarchically from the two circuits of size $N/2$.
As shown in (Figure \ref{fig:hierarchy}, this can be done by interleaving the qubits and inserting the nearest neighbor gates.
Using the hierarchical construction of the sparse coupling graph, which the circuit obeys,
After doubling the system size from $N/2$ to $N$, if the nearest neighbor gates are not present,
then the tripartite mutual information in the linear geometry after $1$ periodic iteration of the circuit is 
$I_{N}(A:B:C)=2I_{N/2}(A:B:C)$,
where $I_{N}(A:B:C)$ represents the tripartite mutual information of a system size $N$. 
The addition of the nearest neighbor gates biases the circuit slightly towards the information to spread locally in the linear geometry.
Therefore, the actual scaling in the linear geometry must be smaller than $2$.

Let $E(N)=N/2(2\log_2 N-1)$ be the number of possible positions which the gates are allowed to be applied 
and $p_0(N)=\frac{2}{2\log_2 N-1}$ to be the probability of the gate application for $s=0$.
On average, the number of non-nearest neighbor gates applied to the circuit is $(E(N)-N)p(N)$, 
which is not enough for having two scrambling circuits of size $N/2$.
However, it is enough for creating one average circuit configuration of the circuit of size $N/2$ at $t=1$.
We can continue this process up to all the system sizes as the number of gates~permits. 

In the large system size limit, the dominant contributor to the tripartite mutual information 
will be the circuit of size $N/2$ and $N/4$.
This is because it is combinatorically unlikely for a set of the left over gates to form a scrambling circuit.
This gives us a recursion~relation:
\begin{align}
   I_N(A:B:C) \sim I_{N/2}(A:B:C)+I_{N/4}(A:B:C)
\end{align}

This relation implies that $I(A:B:C)$ to be monotonically decreasing with respect to the system size provided 
that initial values are negative.

In the limit of large $N$, the recursion relation tells us that $I(A:B:C)$ scales likes a power of the golden ration $\phi$. 
\begin{align}
   I_N(A:B:C) \propto -\phi^{\log_2 N}
\end{align}

\begin{figure}[H]
   \centering
   \includegraphics[width=\textwidth]{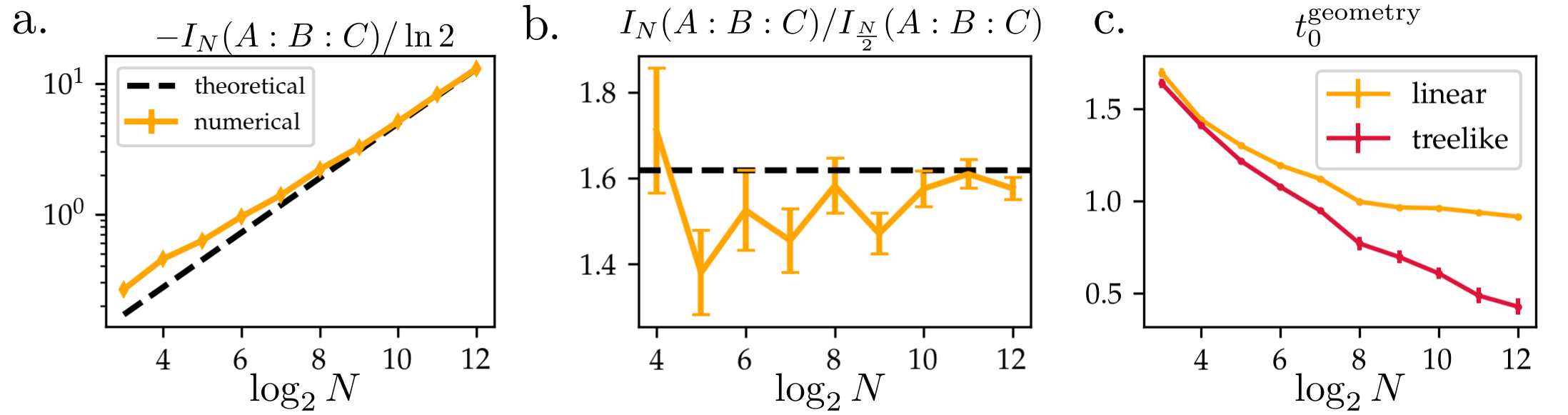}
   \caption{Theoretical behavior of tripartite mutual information at $t=1$ for $s=0$.
      (\textbf{a}) Numerically computed $-I(A:B:C)$ (blue) 
      and theoretically expected values $-I_N(A:B:C)/\ln 2 = I_{2048}\phi^{\log_2 N - 11}/\ln 2$ (red) are plotted
      for the system sizes $N=2^{3},\cdots,2^{12}$.
      The numerically obtained scaling law shows convergence towards the theoretical scaling law as the system size increases.
      (\textbf{b})~The ratio of $I(A:B:C)$ for two different system sizes of $N$ and $N/2$ for different system sizes at $t=1$ and $s=0$
      for the system sizes $N=2^{3},\cdots,2^{12}$.
      The numerically computed values (blue) are converging towards the golden ratio (black dotted) as system size gets larger.
      (\textbf{c}) The $t^{\mathrm{geometry}}_0$ for two different geometries at $s=0$ for $N=2^{3},\cdots,2^{12}$.
      For a sufficiently large $N$, saturation time in the linear geometry is always greater than that of the 
      treelike geometry and hence $t^{\max}_0=t^{\mathrm{linear}}_0$.
      Quantities are calculated by averaging up to $5\times10^{3}$ circuit realizations.
   \label{fig:tpinlimit}}
\end{figure}

Figure \ref{fig:tpinlimit}a shows the theoretical line with $-I(A:B:C)/\ln 2 =  I_{2^{12}}(A:B:C) \phi^{\log_2 N-12}/\ln 2$ 
with a black dotted line and the numerically obtained value of $-I(A:B:C)$ for the different system sizes 
at $t=1$ for $s=0$ (blue line). 
For the large system sizes, the theoretical behavior and the obtained numerical result agrees well. 
Additionally, in Figure~\ref{fig:tpinlimit}b the ratio of tripartite mutual information $I_N(A:B:C)/I_{N/2}(A:B:C)$ at $t=1$ for $s=0$ are plotted.
As expected, the ratio converges towards the golden ratio as the system size increases.
Thus, at $t=1$ and $s=0$, exponentially many more trajectories are expected to have negative tripartite information compare to the 
trajectories with non-negative values; and hence the timescale of $t^{\mathrm{linaer}}_0$ approaches to $1$
in the limit of $N \to \infty$.

The same argument applies for the treelike geometry.
In this case, instead of the nearest-neighbor gates (in the linear geometry), the longest range gates (in the linear geometry) 
has to be taken out.
Due to the circuit structure, which goes from the longest to the shortest range interactions in the treelike geometry,
the time $t^{\mathrm{treelike}}_0$ is smaller than $t^{\mathrm{linear}}_0$ for large $N$ (Figure \ref{fig:tpinlimit}c).
Therefore $t^{\mathrm{linear}}_0$ converging to $t^{\mathrm{linear}}_0=1$ in the limit of large $N$
implies that $t^{\max}_0$ converges towards $t^{\max}_0=1$ constant.
\section[\appendixname~\thesection]{Tripartite Mutual Information at the Scrambling Limit}\label{app:ScramblingLimit}
As shown in \appref{app:stabilizers}, a stabilizer state of $N$ qubits can be represented by a binary matrix $\mathcal{M}$ 
of the dimensions $N$ by $2N$.
A random stabilizer state can therefore be constructed from a random binary matrix with a constraint 
$\mathrm{rank}_{\gft}(\mathcal{M})=N$, where $\mathrm{rank}_{\gft}$ denotes the binary rank of a matrix.
Because the entropy of a subsystem of size $|A|$ is a rank of the corresponding region of the random matrix subtracted by the 
subsystem size, estimation of the scrambled stabilizer state can be calculated by calculating the average binary rank
$\mathrm{rank}_{\gft}(\mathcal{M}_A)$ of a $N \times 2|A|$ binary matrix.

As discussed in \cite{kolchin1998} and Supplementary material D in \cite{hashizumeDeterministicFastScrambling2021},
the probability of a random matrix of size $N$ by $2|A|$ to have a rank of $2|A| - \epsilon$ 
can be written as the following:
\vspace{-9pt}

\begin{adjustwidth}{-\extralength}{0cm}
%\centering %% If there is a figure in wide page, please release command \centering
\centering
\begin{align}
    P\left( \mathrm{rank}_{\gft}(M_{|A|}) = 2|A|-\epsilon \right) &\approx 2^{-\epsilon\left( N-2|A|+\epsilon\right)}
    \times \prod_{i=\epsilon+1}^{\infty} \left( 1-\frac{1}{2^i} \right) 
    \prod_{i=1}^{N-2|A|+\epsilon}\left( 1-\frac{1}{2^{i}} \right)^{-1}.
\end{align}
\end{adjustwidth}
\textls[-15]{where $\epsilon$ is an integer parameterizing how far the matrix is from full rank. The average entropy of a subsystem of size $|A|$ of a random stabilizer state is then, for large $N$, approximately,}
\begin{align}
    \langle  S_{A} \rangle 
    \approx \left(|A| - 2^{2|A| - N}\right)\ln 2.
\end{align}

With this, the tripartite mutual information in the scrambling limit becomes
\begin{align}
   I(A:B:C) \approx -S_{BC} \approx -\frac{N}{2}\ln 2.
\end{align}

\section[\appendixname~\thesection]{Time Dependence of Teleportation Fidelities}
\label{app:TDTF}
\subsection{Teleportation Fidelity for Fixed $s$ and Varying Sites $B$}
\label{app:TDTF_fixed_s}

\begin{figure}[H]
   \includegraphics[width=3in]{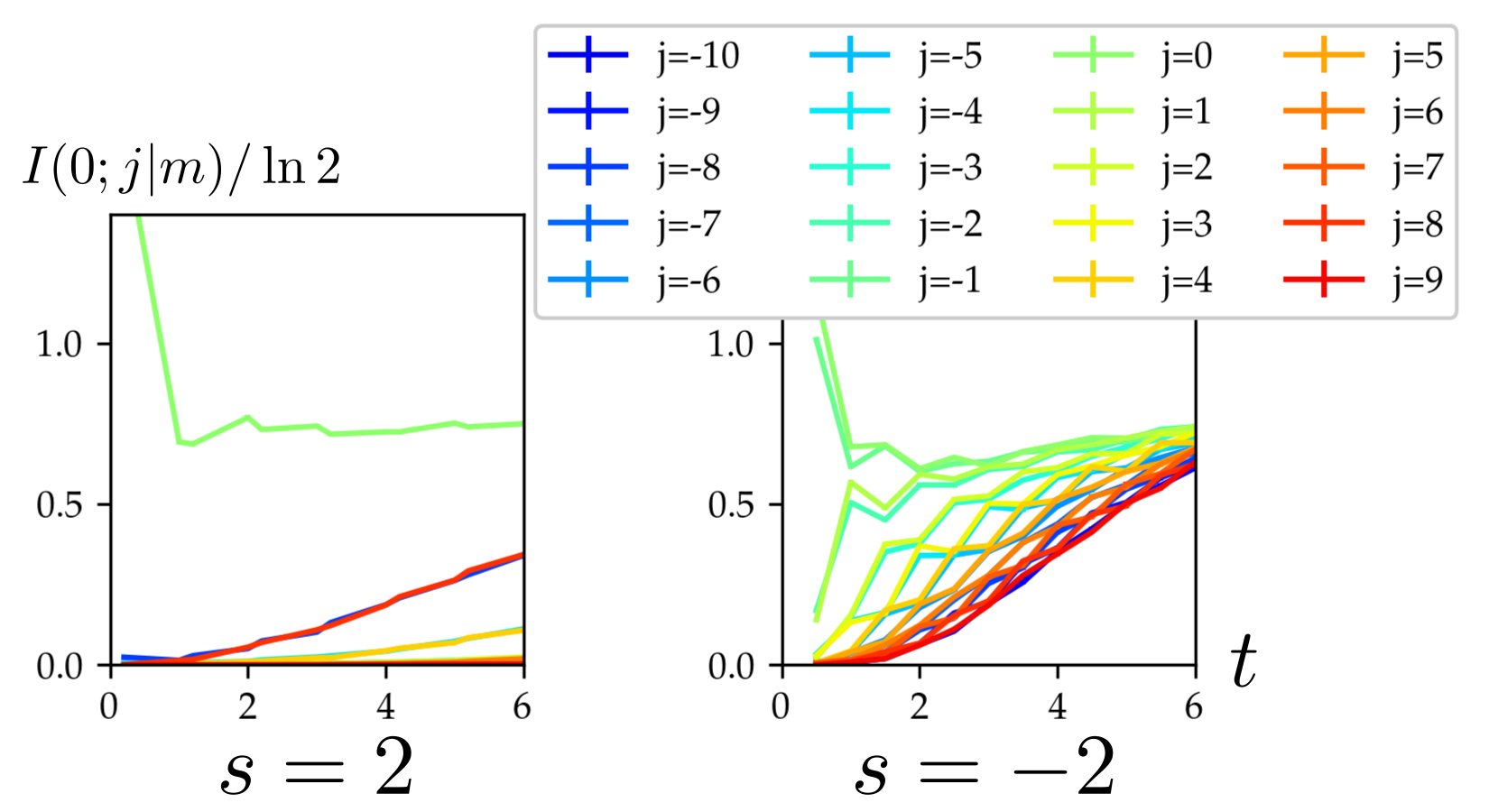}
   \caption{
      Time dependence of teleportation fidelity for different sites.
   The time dependence of the teleportation fidelity $I(0;j \rvert m)$ 
   evolved up to $t=6$ with system size $N=128$ for $s=-2$ and $s=2$ for $j=-10,-9,\cdots,8,\,9$.
   The mean values are calculated and the errors are estimated from up to $1.5\times10^{4}$ trajectories.
   \label{fig:ATeleport_s_error}}
\end{figure}

The color plot in Figure \ref{fig:telelightcone}a is a plot of teleportation fidelity $I(0;j \rvert m)$ of a state evolved 
in the sparse Clifford circuit with system size $N=128$.
It is plotted for $s=-2$ (left) and $s=2$ (right) for the sites $j=-10,-9,\cdots,8,9$.
In Figure \ref{fig:ATeleport_s_error}, we show the time dependency of the quantity from $t=0$ to $t=6$
for $s=2$ (left) and $s=-2$ (right) at the different sites (the lines of various colors). 
The mean values are obtained and errors are estimated from $1.5\times10^{4}$ trajectories.

\subsection{Teleportation Fidelity for Fixed Sites and Varying $s$}

\begin{figure}[H]
   \centering
   \includegraphics[width=\textwidth]{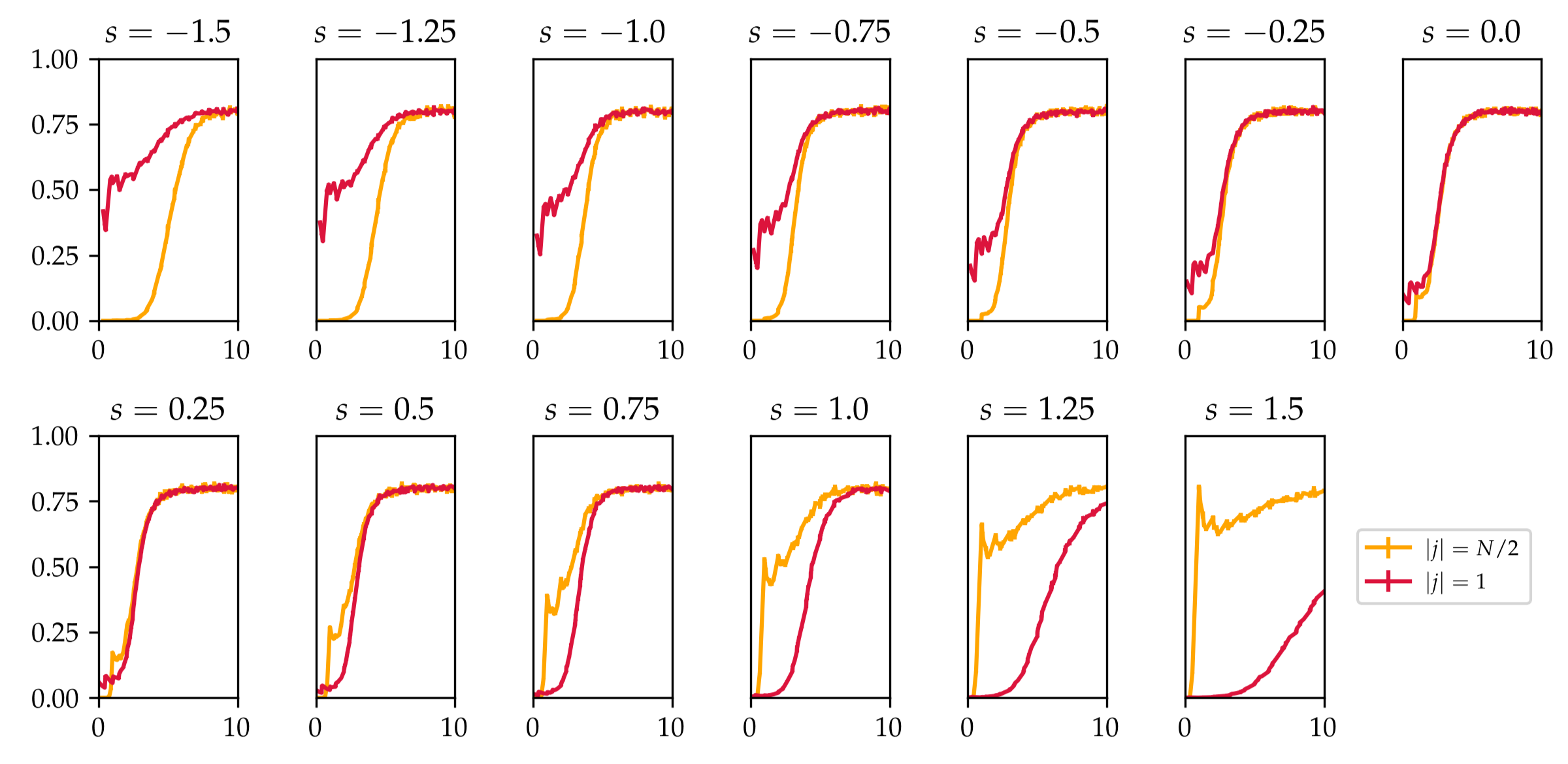}
   \caption{
      Time dependence of teleportation fidelity for fixed $s$.
   The time dependence of the teleportation fidelity $I(0;j\rvert m)$ 
   evolved up to $t=6$ with system size $N=128$ for 
   $s=-1.5,-1.25,-1,-0.75,-0.5,-0.25,0,0.25,0.5,0.75,1,1.25,1.5$.
   The teleportation fidelity is calculated for $|j|=N/2$ for characterizing the linear (Euclidean) geometry
   and the average of $j=1$ and $j=-1$ for characterizing the treelike (Ultrametric) geometry.
   The mean values are calculated and the errors are estimated from up to $1.5\times10^{4}$ trajectories.
   \label{fig:ATeleport_B_error}}
\end{figure}

The color plot in Figure \ref{fig:telelightcone}c is a plot of teleportation fidelity $I(0;j \rvert m)$ of a state evolved 
in the sparse Clifford circuit with system size $N=128$.
It is plotted for $j=N/2$ and the average of $j=-1$ and $j=1$ to characterize the behavior of the teleportation fidelity in 
linear and treelike, respectively, for $s=-1.5,-1.25,-1,-0.75,-0.5,-0.25,0,0.25,0.5,0.75,1,1.25,1.5$.
In Figure \ref{fig:ATeleport_B_error}, we show the time dependence of the quantity from $t=0$ to $t=10$ 
for the values of $s$ mentioned above, for $j=N/2$ (orange) and average of $j=-1$ and $j=1$ (red).
The mean values are obtained and errors are estimated from $1.5\times10^{4}$ trajectories. 

\section[\appendixname~\thesection]{Finite Size Scaling for Finding $\mathbf{\emph{t}}_\mathbf{\emph{c}}$ of Teleportation Fidelity}
\label{app:FSScaling}

\begin{figure}[H]
   \centering
   \includegraphics[width=\textwidth]{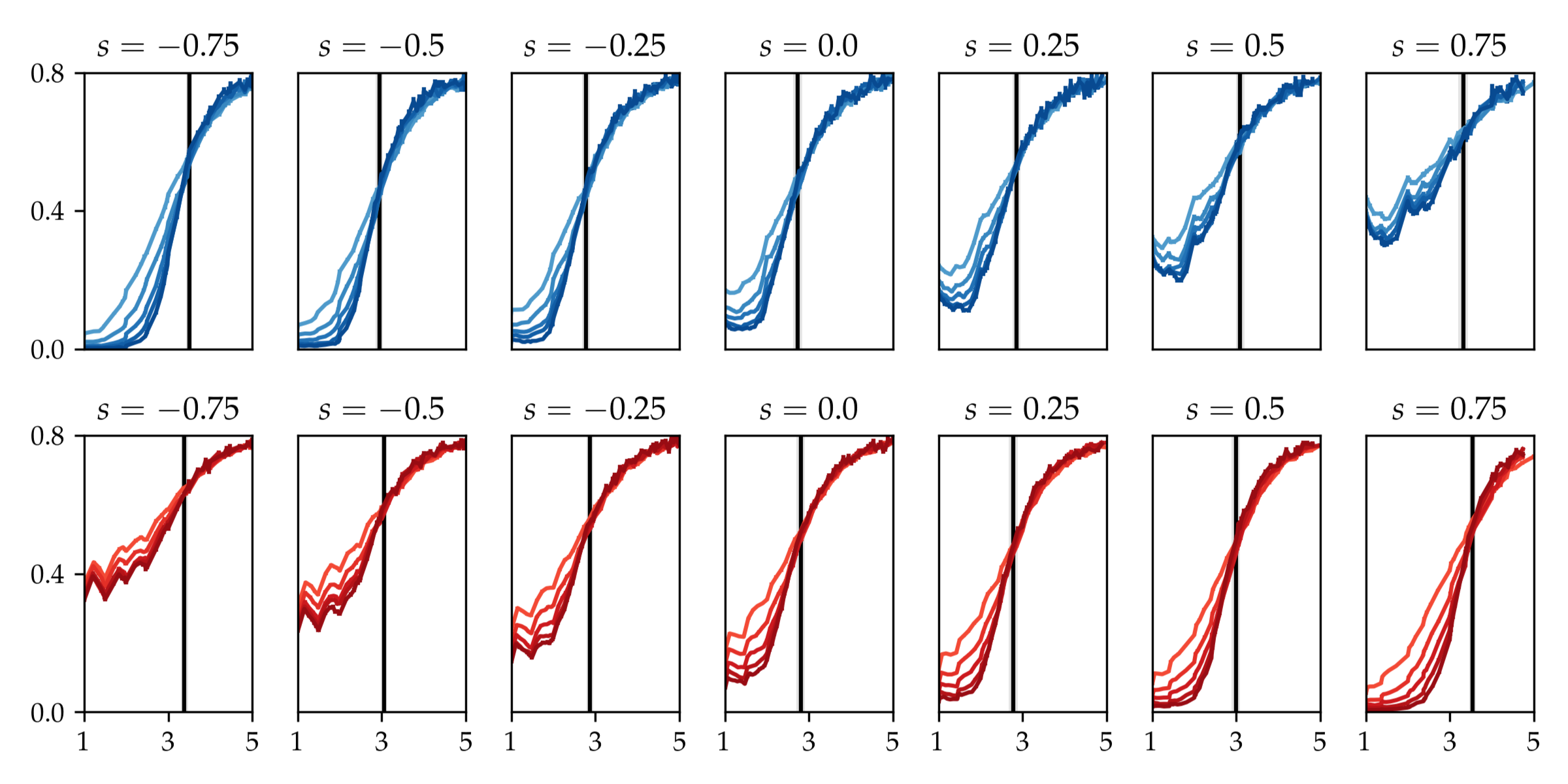}
   \caption{
      Time dependence of teleportation fidelity for different system sizes.
   The time dependence of the teleportation fidelity $I(0;j \rvert m)$ evolved up to $t=5$ for 
   $s=-0.75,-0.5,-0.25,0,0.25,0.5,0.75$ for the system sizes $N=2^{5},2^{6},\cdots,2^{9}$ (light to dark).
   The quantity is calculated for the sites $j=N/2$ for characterizing the linear (Euclidean) geometry (upper panels, blue) 
   and average between $j=-1$ and $j=1$ for characterizing the treelike geometry (lower panels, red).
   The mean values are calculated and the errors are estimated from up to $1.5\times10^{4}$ trajectories.
   \label{fig:ATeleport_fs}}
\end{figure}

The critical time $t_c$ which divides the non-teleporting $t<t_c$ and teleporting $t_c<t$ regimes
shown in the black line in Figure \ref{fig:telelightcone}c is calculated from the crossing points of the teleportation fidelity 
for the different system sizes.
Here, in Figure \ref{fig:ATeleport_fs}, the quantity is plotted for the values of $s=-0.75,-0.5,-0.25,0,0.25,0.5,0.75$
for the system sizes $N=2^{5},2^{6},\cdots,2^{9}$ (light to dark).
The quantity is calculated for the sites $j=N/2$ for characterizing the linear (Euclidean) geometry (upper panels, blue) 
and average between $j=-1$ and $j=1$ for characterizing the treelike geometry (lower panels, red).
The mean values are calculated and the errors are estimated from up to $1.5\times10^{4}$ trajectories.
Crossings are not observed for $s\leq-1.0$ and $1.0\leq s$ in the system sizes that we investigated.

%%%%%%%%%%%%%%%%%%%%%%%%%%%%%%%%%%%%%%%%%%
%}}}

%{{{ References
\begin{adjustwidth}{-\extralength}{0cm}
\reftitle{References}
\bibliography{References}
\end{adjustwidth}
\end{document}